\documentclass[final,10pt,3p]{elsarticle}

\journal{Astronomical Journal}

\usepackage{amssymb}
\usepackage{amsthm}
\usepackage{amsmath}
\usepackage{hyperref}
\hypersetup{
    colorlinks=true,
    linkcolor=blue,
    filecolor=magenta,      
    urlcolor=cyan,
}

\usepackage[nameinlink,noabbrev]{cleveref}

\usepackage{longtable}
\usepackage{setspace}
\usepackage{float}
\floatstyle{plaintop}
\restylefloat{table}

\usepackage{ltcaption}

\usepackage{subfigure}
\usepackage{graphicx}

\usepackage{xfrac}

\usepackage{lineno}

\usepackage{color, colortbl}
\definecolor{Gray}{gray}{0.9}

\biboptions{authoryear,round}

\begin{document}

\begin{frontmatter}

\title{Thermophysical Modeling of Asteroid Surfaces using Ellipsoid Shape Models}

\author[label1]{Eric M. MacLennan\corref{cor1}}
\address[label1]{Earth and Planetary Sciences Department, Planetary Geosciences Institute, The University of Tennessee, Knoxville, TN 37996}

\cortext[cor1]{corresponding author}

\ead{emaclenn@vols.utk.edu}
\ead[url]{web.utk.edu/~emaclenn/home}

\author[label1]{Joshua P. Emery}

\begin{abstract}
Thermophysical Models (TPMs), which have proven to be a powerful tool in the interpretation of the infrared emission of asteroid surfaces, typically make use of {\it a priori} obtained shape models and spin axes for use as input boundary conditions. We test then employ a TPM approach - under an assumption of an ellipsoidal shape - that exploits the combination of thermal multi-wavelength observations obtained at pre- and post-opposition. Thermal infrared data, when available, at these observing circumstances are inherently advantageous in constraining thermal inertia and sense of spin, among other physical traits. We show that, despite the lack of {\it a priori} knowledge mentioned above, the size, albedo, and thermal inertia of an object are well-constrained with precision comparable to that of previous techniques. Useful estimates of the surface roughness, shape, and spin direction can also be made, to varying degrees of success. Applying the method to WISE observations, we present best-fit size, albedo, thermal inertia, surface roughness, shape elongation and sense of spin direction for 21 asteroids. We explore the thermal inertia's correlation with asteroid diameter, after accounting for its dependence on the heliocentric distance.
\end{abstract}


\end{frontmatter}


\section{Introduction}\label{sec:1}
Multi-wavelength, photometric infrared observations of asteroids provide essential information about the thermophyiscal properties of their surfaces. A handful of both simple thermal models \citep{Harris98,Lebofsky_etal78,Lebofsky_etal86,Wolters&Green09,Myhrvold16} and more sophisticated themophysical models \citep[TPMs;][]{Spencer_etal89,Spencer90,Lagerros96,Delbo_etal07,Mueller07,Rozitis&Green11} have been established as effective means of modeling thermal infrared observations of asteroids. The general purpose of a thermal model is to compute surface temperatures for an object, which are in turn used to calculate the emitted flux at the desired wavelengths \citep[see][for a recent review]{Delbo_etal15}. Thermophysical models have proved to be a powerful tool in providing meaningful estimates of an object's size and albedo and provide insight into thermophysical characteristics of asteroid regoliths \citep{Emery_etal14,Hanus_etal15,Hanus_etal18,Landsman_etal18,Rozitis&Green14,Rozitis_etal14,Rozitis_etal18}.

Most simple thermal models assume an idealized (often spherical) object shape, instead of including {\it a priori} knowledge of the shape, in order to estimate the diameter and albedo of an object. However, unlike TPMs, simple thermal models lack the ability to estimate geologically-relevant thermophysical properties such as the thermal inertia. TPMs require a spin vector - typically sourced from the Database of Asteroid Models from Inversion Techniques \citep[DAMIT\footnote{\url{http://astro.troja.mff.cuni.cz/projects/asteroids3D/web.php}};][]{Durech_etal10} - and object shape model as input \citep[e.g.,][]{Hanus_etal18}.

With the recent surge in thermal infrared observations from large-scale surveys, an ever-increasing number of asteroids are being observed across several epochs, building up sets of observations that span both pre- and post-opposition and often at large solar phase angles \citep{Mainzer_etal11a}. Such observations provide additional constraints and/or allow for more free parameters in data-modeling inversion techniques. However, the rate of thermal infrared observations significantly outpace the efforts to characterize the shapes of individual asteroids. Advantages of multi-epoch observations have previously been noted \citep[e.g.][]{Spencer90} and been used recently to derive thermophysical and spin properties of objects \citep{Muller_etal11,Muller_etal14,Muller_etal17,Durech_etal17}. These studies serve as the foundation on which we build a methodological approach aimed at extracting important thermophysical properties from the large number of asteroids observed at pre- and post-opposition at multiple thermal wavelengths often acquired from thermal infrared surveys.

In this paper, we outline and test a TPM approach that utilizes pre- and post-opposition multi-wavelength thermal observations when information about an object is limited. To do so, we use various simple shapes (both spherical and prolate ellipsoids) of varying spin vectors. The goal of this work is to demonstrate and establish the effectiveness of this modeling approach and compare it to previous studies. In \cref{sec2} we briefly review established thermal modeling techniques and their ability to constrain diameter, albedo, thermal inertia, surface roughness, shape and spin direction - highlighting the use of pre-/post- opposition observing geometries. We describe our thermophysical model and its implementation in \cref{sec3}. In \cref{sec4} we implement and analyze the effectiveness of our approach to that of the previously reviewed works. We then apply our modeling approach and present results for 21 asteroids in \cref{sec5}. In a follow-up paper, we will supplement the number of objects analyzed with this method by an order of magnitude.


\section{Thermal Modeling Background and Motivation}\label{sec2}

A few simple thermal models and TPMs have been developed to estimate various physical and thermophyphysical properties of asteroids. Many thermal models attempt to match a disk-integrated flux to a set of telescopic observational data by calculating surface temperatures for a given shape. Simple thermal models model surface temperatures by using closed-form equations (which can be evaluated in a finite number of operations) based off of the equilibrium surface temperature ($T_{eq}$), which is calculated via the energy balance between the amount of absorbed insolation and emitted thermal energy:
\begin{equation}\label{eq1}
  \frac{S_{\odot}(1-A)}{R^{2}_{AU}} - \varepsilon_B \sigma T^{4}_{eq} = 0.
\end{equation}
In \cref{eq1}, $S_{\odot}$ is the solar constant at 1 AU \citep[1367 Wm$^{-2}$; ][]{Frohlich09}, $A$ is the bolometric Bond albedo, $R_{AU}$ is the heliocentric distance in Astronomical Units, $\varepsilon_B$ is the bolometric emissivity, and $\sigma$ is the Stefan-Boltzmann constant. The ``beaming parameter'', $\eta$, modifies the energy balance in order to account for various thermophysical effects that cause temperatures to depart from thermal equilibrium.

On the other hand, TPMs compute surface temperatures using Fourier's Law of heat-diffusion (evaluated numerically). Well established simple thermal models include the Standard Thermal Model \citep[STM;][and references within]{Lebofsky_etal86}, Near-Earth Asteroid Thermal Model \citep[NEATM;][]{Harris98}, and the Fast Rotating Model \citep[FRM;][]{Lebofsky_etal78} More recently, the Night Emission Simulated Thermal Model \citep[NESTM;][]{Wolters&Green09}) and Generalized FRM \citep[GFRM;][]{Myhrvold16} have been created by modifying the NEATM and FRM, respectively. For reasons of applicability and flexibility, especially when analyzing large datasets, the NEATM has been used most often \citep{Trilling_etal10,Mainzer_etal11b,Masiero_etal11}. A brief description and history of the STM, FRM, and NEATM can be found in \cite{Harris&Lagerros02}. The NEATM has been the preferred simple thermal model for multi-wavelength thermal surveys; due to the ease of application to single epoch observations of objects for which no shape or spin information exists. But, a TPM is the preferred tool for interpreting the multi-wavelength thermal observations of objects with shape and spin vector information.

\subsection{Thermophysical Models}\label{subsec:22}

Many versions of TPMs, with varying levels of complexity, have been developed throughout the past few decades. Such models explicitly account for the effects that physical attributes of the surface have on the thermal emission \citep{Spencer_etal89}. Subsurface heat conduction, self-shadowing from direct insolation (incoming solar radiation), multiply scattered insolation, and re-absorbed thermal radiation (i.e. self-heating) are implemented in a multitude of ways. The energy conservation expressed by \cref{eq1} is amended to include additional terms that account for these effects:
\begin{equation}\label{eq3}
  \frac{S_{\odot}(1-A)}{R^{2}_{AU}}\cos(i)(1-s) + E^\mathit{solar}_\mathit{scat} + E^\mathit{therm}_\mathit{abs} + k \frac{dT}{dx} \bigg|_\mathit{surf} - \varepsilon_B \sigma T^{4}_\mathit{surf} = 0.
\end{equation}
Where, $k$ is the {\it effective} thermal conductivity, $i$ is the angle between a surface facet's local zenith and sun-direction (solar incidence angle), and $s$ is a binary factor indicating whether an element is shadowed by another facet ($s=1$), or not ($s=0$). The terms $E^\mathit{solar}_\mathit{scat}$ and $E^\mathit{therm}_\mathit{abs}$ encompass the contributed energy input from scattered solar radiation and re-radiated thermal photons, respectively, from other surface elements. Rough topography has been modeled as spherical section craters \citep{Spencer90,Emery_etal98}, random Gaussian surface \citep{Rozitis&Green11}, and using fractal geometry \citep{Davidsson&Rickman14}. The mathematics and numerical implementations of these features differ somewhat from one another. For further details, we refer the reader to the primary papers and to \cite{Davidsson_etal15} for a comparison of the different implementations.

TPMs often use the time-dependent, one-dimensional heat diffusion equation,
\begin{equation}\label{eq4}
  \frac{\partial T(x,t)}{\partial t} = \frac{k}{\rho c} \frac{\partial^{2} T(x,t)}{\partial x^{2}},
\end{equation}
to model the flow of heat into and out from the subsurface. This presentation of Fourier's Law in \cref{eq4} assumes that the {\it effective} thermophysical factors (thermal conductivity, bulk density, $\rho$, and specific heat capacity, $c$) do not vary with depth or temperature. One-directional heat flow into the subsurface is also often assumed, but is well-justified by the concept that thermal energy flow is aligned with the temperature gradient - either directly up or down beneath the surface.

In the subsections below, we review and assess how various physical parameters (diameter, albedo, thermal inertia, surface roughness, shape, and spin direction) can be constrained in various observing circumstances. In particular, we highlight the optimal observing configurations and possible biases that may arise from particular viewing geometries.

\subsection{Diameter and Albedo}\label{subsec:23}

One of the primary motivations for developing thermal models was to obtain size estimates of objects from disk-integrated thermal infrared observations \citep[e.g.][]{Allen70, Morrison73}, since thermal flux emission is directly proportional to the object's projected area (i.e. \cref{eq9}). A single value of diameter is not uniquely defined for a non-spherical body, so an {\it effective diameter} ($D_{\mathit{eff}}$) is given: the diameter of the sphere \citep{Mueller07} having the same projected area as the object. During the span of data collection for a non-spherical body, the projected area will almost never be constant. In this case, $D_\mathit{eff}$ is best reported as a time-averaged value, or adjusted using visible lightcurve data obtained simultaneously or proximal in time to the thermal observations \citep[e.g.][]{Lebofsky&Rieke79,Harris&Davies99,Delbo_etal03,Lim_etal11}. If a detailed shape model is being used, then the rotational phase of the object can be shifted in order to match the time-varying flux \citep{AliLagoa_etal13}.

Geometric albedo ($p_{V}$) is calculated directly from $D_{\mathit{eff}}$ and the object's absolute magnitude ($H_{V}$) \citep{Russell1916, Pravec&Harris07}:
\begin{equation}\label{eq12}
\sqrt{p_{V}} = \dfrac{1329\ \textrm{km}\ 10^{-0.2H_{V}}}{D_{\mathit{eff}} [\textrm{km}]}
\end{equation}
The geometric albedo is then converted to the bolometric Bond albedo using the definition of the phase integral, $q$:
\begin{equation}\label{eq13}
  q \equiv A_{V}/p_{V} = 0.290 + 0.684 \times G_{V}
\end{equation}
in which $G$ is the slope parameter \citep{Bowell_etal89} and the approximation $A \approx A_{V}$ is made in \cref{eq1}.

\subsection{Thermal Inertia and Surface Roughness}\label{subsec:24}

\begin{figure}[H]
  \centering
  \begin{tabular}[b]{l}
	\includegraphics[clip,trim = 0.6cm 0.25cm 1cm 1cm,width=.45\linewidth]{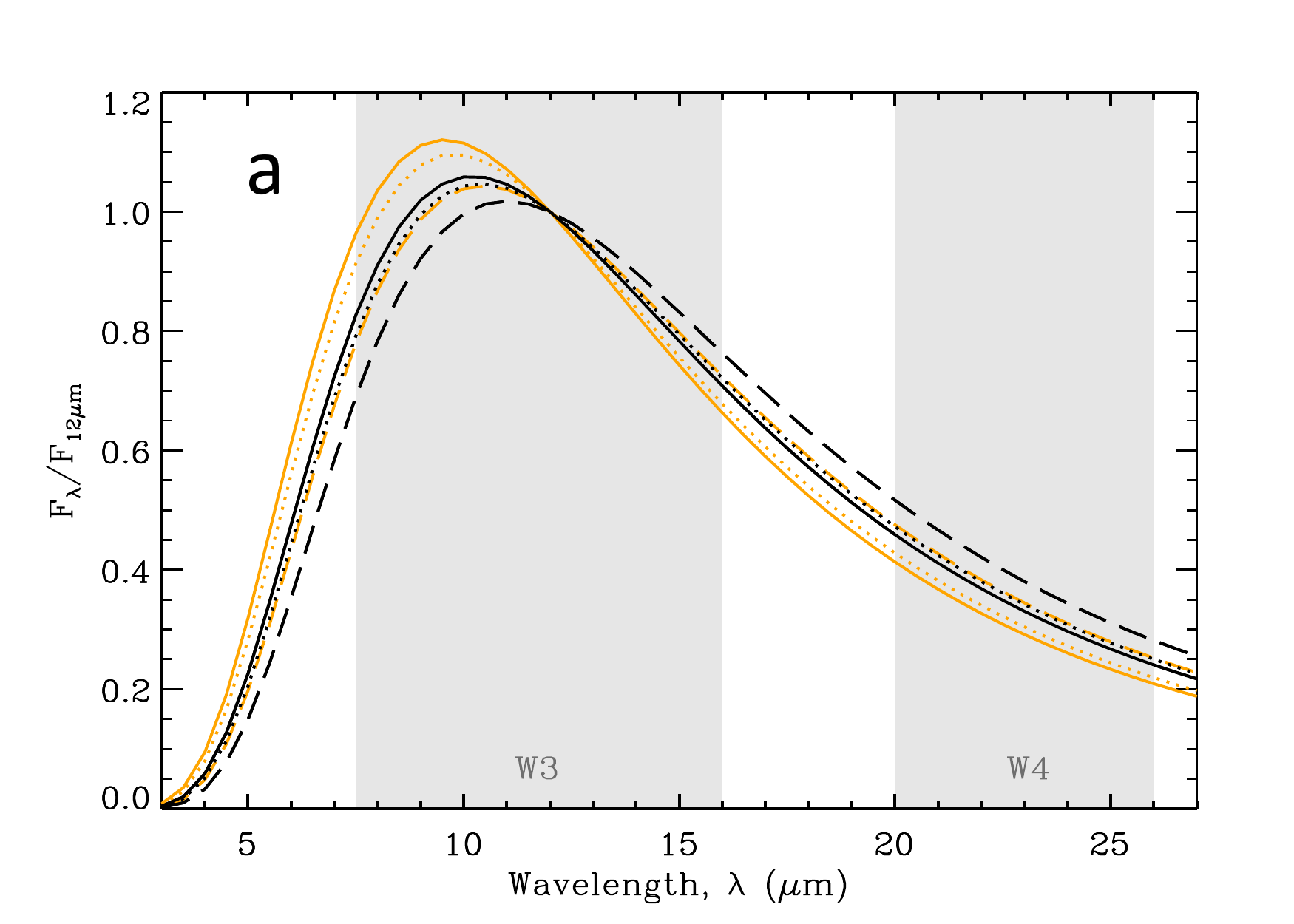}
	\includegraphics[clip,trim = 0.6cm 0.25cm 1cm 1cm,width=.45\linewidth]{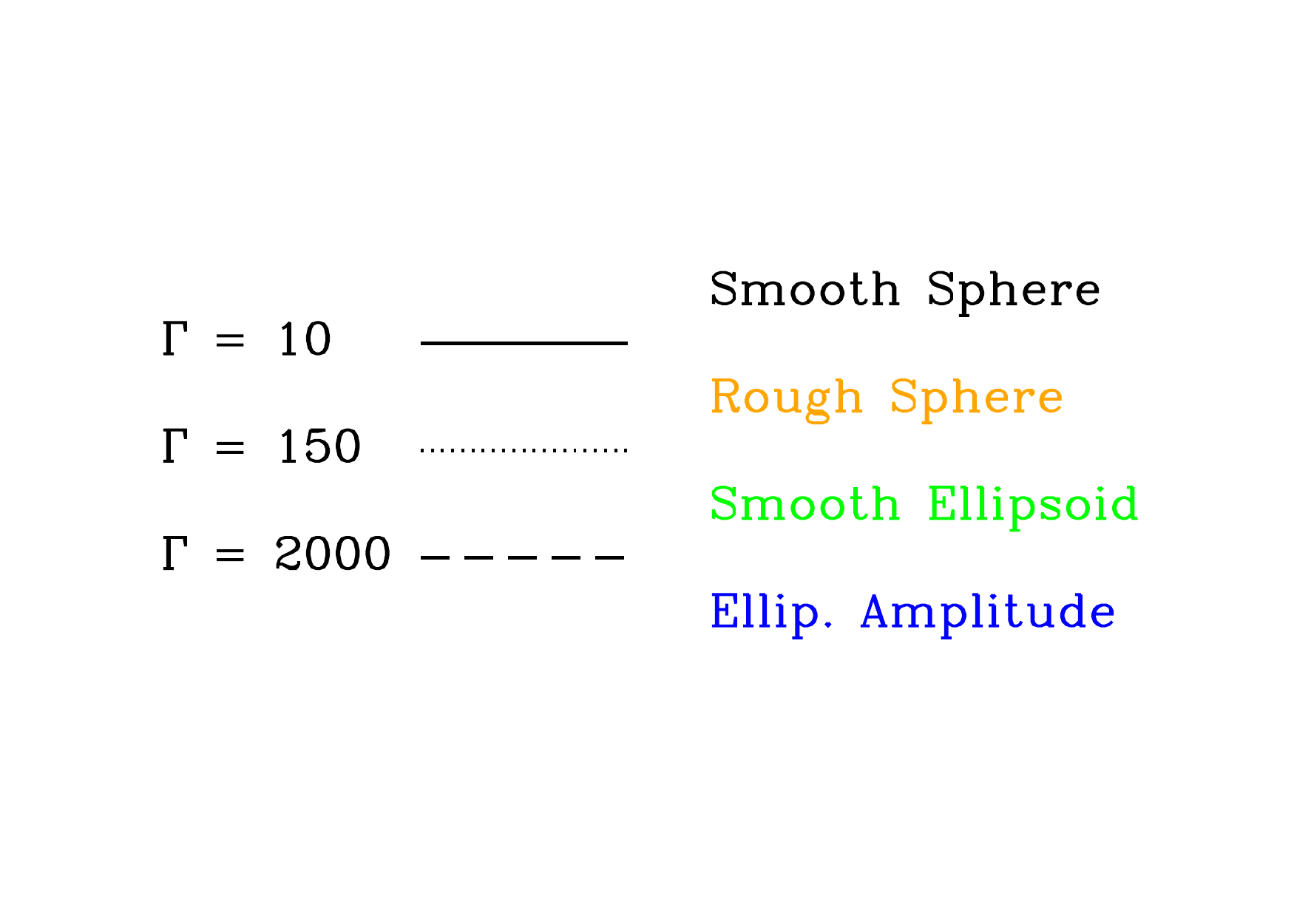} \\
	\includegraphics[clip,trim = 0.6cm 0.25cm 1cm 1cm,width=.45\linewidth]{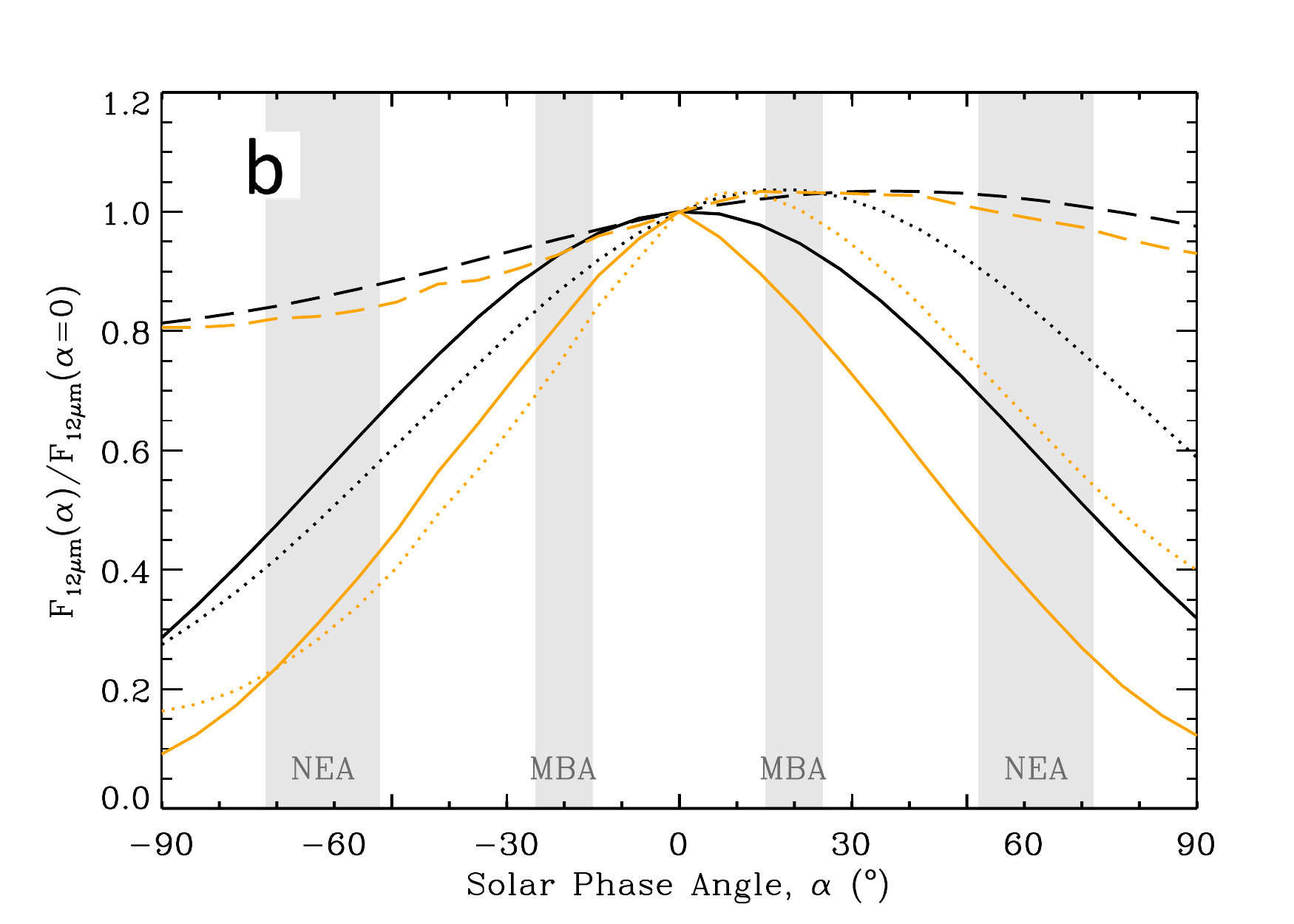}
	\includegraphics[clip,trim = 0.6cm 0.25cm 1cm 1cm,width=.45\linewidth]{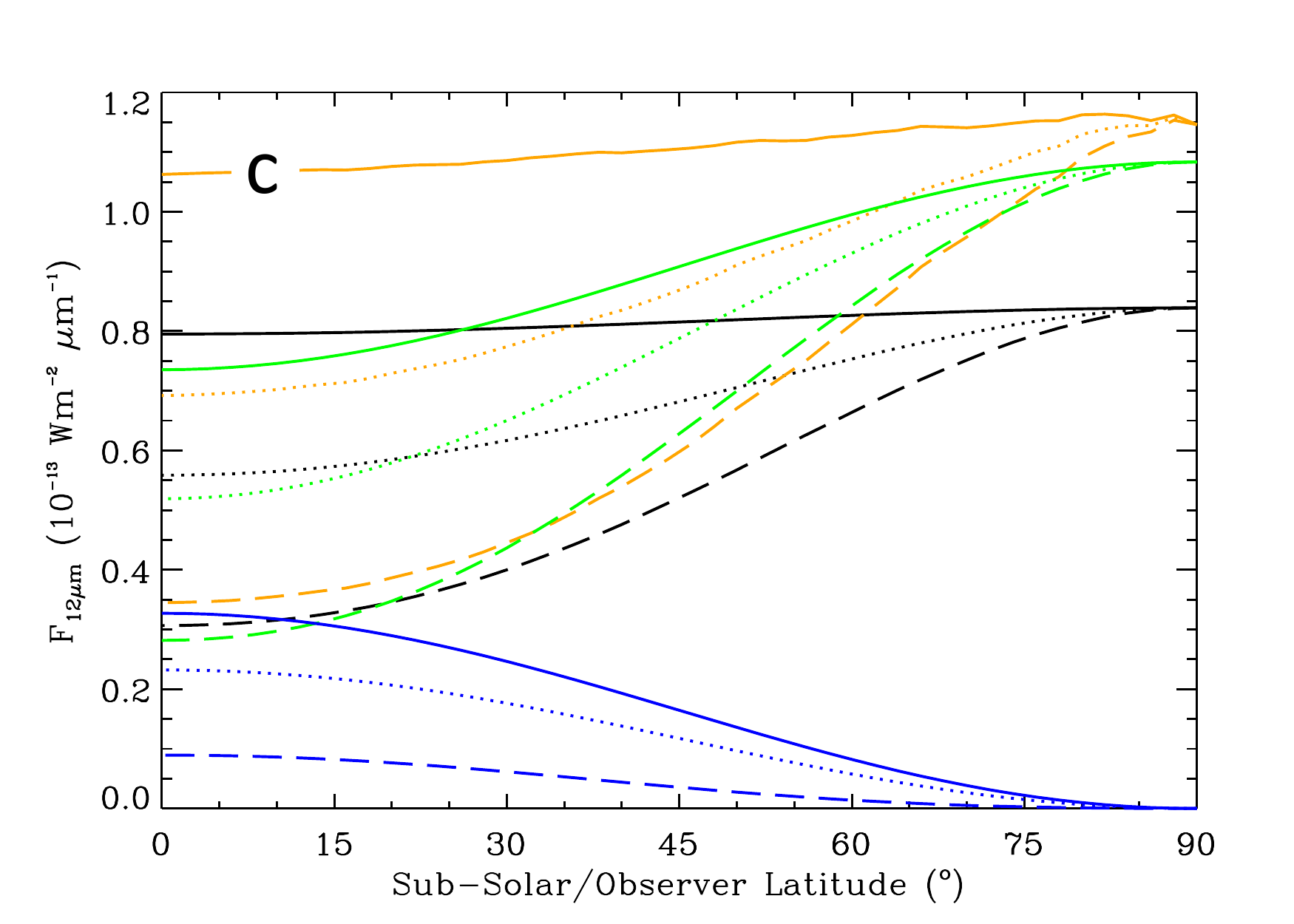} \\
  \end{tabular}
  \caption{Comparison of thermal flux emitted from an object varying it's thermophysical properties and shape, as a computed from the TPM described in \cref{sec3}. Panel (a) shows blackbody curves, normalized at 12 $\mu$m, thermal phase curves are shown in (b), and panel (c) shows the flux as a function of the sub-solar/observer latitude. The modeled object has $D_{\mathit{eff}} = 1$ km, $T_{eq} = 300$ K, $P_\mathit{rot} = 10$ hr, located $1$ AU from the observer. Within all frames, black curves are for a smooth sphere, orange curves are for a rough ($\bar{\theta} = 29^{\circ}$) sphere, and green curves are for a prolate ellipsoid with $a/b = 1.75$. Blue curves show the amplitude of the ellipsoid's thermal lightcurve. As indicated in the key, the solid, dotted, and dashed curves in all frames distinguish different values of thermal inertia. We note here that pre-opposition is defined as $\alpha > 0$, in which the afternoon side of a prograde rotator is viewed.}\label{fig1}
\end{figure}

The characteristic ability of a material to resist temperature changes when subject to a change in energy balance is quantified by its thermal inertia ($\Gamma = \sqrt{k\rho c}$). Atmosphereless surfaces with low thermal inertia conduct little thermal energy into the subsurface, resulting in dayside temperatures that are close to instantaneous equilibrium with the insolation and extremely high diurnal temperature differences. Surfaces having large thermal inertia store and/or conduct thermal energy in the subsurface during the day so that significant energy is re-radiated during the nighttime hours, which results in a comparatively smaller diurnal temperature change. This influence that thermal inertia has on the surface temperature distribution manifests in the observed SED. Thus photometry at two wavelengths (e.g. a measurement of the color temperature) can provide a better measure of thermal inertia compared to that of a single wavelength \citep{Mueller07}. \Cref{fig1}(a) shows the normalized SEDs of surfaces (at opposition) having different thermal inertia, roughness, and shape. Surfaces with higher roughness and low thermal inertia exhibit flux enhancement at small wavelengths near opposition, lowering the wavelength location of the peak flux. Data collected which have large wavelength separation that span the blackbody peak are most useful, since they are sensitive to relatively warmer and cooler portions of the surface - in other words the overall temperature distribution. \cite{Delbo&Tanga09} estimated thermal inertias for 10 main-belt asteroids using multi-wavelength IRAS (Infrared Astronomical Satellite) observations, and \cite{Hanus_etal15,Hanus_etal18} collectively present thermal inertia estimates for 131 objects using WISE (Wide-Field infrared Survey Explorer) data. \cite{Harris&Drube16} present population trends using over a hundred thermal inertias derived from the NEATM $\eta$ values of asteroid observed with WISE.

Multiple observations at a single wavelength but at various solar phase angles - a {\it thermal phase curve} - have also proven to be useful methods to estimate thermal inertia \citep[e.g.][]{Spencer90}. \Cref{fig1}(b) shows examples of thermal phase curves for 3 different objects, each possessing 5 different thermal inertia values. \cite{Muller_etal11} and \cite{Muller_etal17} demonstrated that observations of (162173) 1993 $\textrm{JU}_3$, Ryugu, taken on one side of opposition but widely spaced in solar phase angle ($30^{\circ}$) can constrain thermal inertia as they effectively had two points along a thermal phase curve. As seen in \cref{fig1}(b), this approach can be optimized when the thermal phase curve spans across both sides of opposition. Thus, observing at pre- and post-opposition virtually guarantees that thermal emission information from the warmer afternoon ($\alpha > 0$) and cooler morning sides is gathered \citep{Muller_etal14}.

Surfaces with large degrees of roughness exhibit warmer dayside temperatures, due to more of the surface area pointed towards the sun and from the effects of multiple scattering of reflected and emitted light. The term ``beaming effect'' is used to describe enhanced thermal flux return in the direction of the sun at low phase angles. As pointed out by \cite{Rozitis17}, the flux enhancement near opposition is highly sensitive to surface roughness. They also demonstrated that telescopic observations at a near pole-on illumination and viewing geometry can be used to effectively constrain the degree of roughness, specifically when $\alpha < 40^{\circ}$ and the sub-solar latitude is greater than $60^{\circ}$ (\cref{fig1}(b \& c)). \Cref{fig1}(b) shows thermal phase curves of objects with smooth and rough surfaces of varying thermal inertia. \Cref{fig1}(c) shows the thermal flux emitted as a function of sub-solar latitude and recreates the findings of \cite{Rozitis17}. In general, the surface roughness of airless bodies are more easily estimated near opposition or, better yet, with disk-resolved observations that offer a greater range of viewing geometries. The \cite{Rozitis17} study shows that the uncertainty in thermal inertia is slightly larger for an object observed at a large phase angle, as a consequence of the difficulty in estimating surface roughness.

\subsection{Shape and Spin Direction}\label{subsec:25}

Disk-integrated flux is also directly affected by the shape and spin vector of an object \citep{Durech_etal17}. If thermal observations sample the entire rotation period, a thermal light curve is useful to constraining these parameters. In particular, the amplitude of the thermal lightcurve is largely affected by the elongation and orientation of the spin vector. In principle, constraints can be placed on the spin vector - or, at least, spin {\it direction} - of objects since the temperature distribution is affected by the sub-solar latitude, and thus the spin vector \citep{Muller_etal11,Muller_etal12,Muller_etal14,Muller_etal17}. \Cref{fig1}(c) shows the variation of the flux for an ellipsoid at different viewing geometries. Gathering multiple thermal lightcurves can offer amplitude measurements at different viewing geometries, which significantly improves determination of the global shape. Both \cite{Morrison77} and \cite{Hansen77b} correctly determine the spin direction by observing the change in diameter estimates for data acquired before and after opposition. The diameter estimate of (1) Ceres, a prograde rotator, before opposition was $5\%$ larger than the diameter estimates from post-opposition observations. \cite{Muller_etal14} explicitly mentions that observations at either side of opposition are useful indicators of the spin direction. As shown in \cref{fig1}(a), this effect is due to the flux excess emitted from the hotter afternoon side of a prograde rotator as it is would be observed before opposition \citep[e.g.][]{Lagerros96}.

\subsection{Motivation for this Work}

The text and figure above demonstrate how multi-wavelength thermal lightcurve observations, which span across an object's blackbody peak region, and at both pre- and post-opposition (with $\Delta \alpha > 40^{\circ}$), contain information directly dependent on an object's physical properties. Thermal inertia, surface roughness, and spin direction can all be constrained from measurements of an object's thermal phase curve, particularly when observations widely vary in solar phase angle. Additionally, thermal lightcurve amplitude measurements at more than one viewing geometry are a unique indicator of the elongation of a rotating object. The work described in the rest of this paper demonstrates and quantifies the effectiveness of using an ellipsoidal TPM to model the thermophysical properties of any given asteroid, given the dataset described above.


\section{Thermophysical Model Description}\label{sec3}

The TPM approach described here involves calculating surface temperatures for a spherical and various prolate ellipsoids in order to model and fit  pre- and post-opposition multi-wavelength thermal observations. Both a smooth and rough surface TPM were developed for use in the fitting routine.

\subsection{Smooth TPM}\label{subsec:31}

In development of the smooth TPM we find it particularly useful to parameterize the depth variable in \cref{eq4} as $x'=x/l_{s}$ \citep{Spencer_etal89}, where $l_{s}$ is the thermal skin depth, the length scale at which the amplitude of the diurnal temperature variation changes by a factor of $e\approx2.718$:
\begin{equation}
	l_{s} = \sqrt{\frac{k}{\rho c}\frac{P_\mathit{rot}}{2\pi}},
\end{equation}
where $P_\mathit{rot}$ is the rotation period of the body. This parameterization transforms the temperature gradient term in \cref{eq3} as follows:
\begin{equation}
	k \frac{dT}{dx'} \bigg|_\mathit{surf} \Rightarrow \Gamma \sqrt{\frac{2\pi}{P_\mathit{rot}}} \frac{dT}{dx'} \bigg|_\mathit{surf},
\end{equation}
In order to further reduce the number of independent input variables, we also parameterize the temperature and time as $T'=T/T_{eq}$ and $t'=2\pi t/P_\mathit{rot}\Theta$. The thermal parameter, $\Theta$\footnote{For reference, objects with $\Theta=0$ exhibit surface temperatures in equilibrium with the insolation and objects with $\Theta>100$ have nearly isothermal surface temperatures at each latitude.} is given by:
\begin{equation}\label{eq14}
	\Theta = \frac{\Gamma}{\varepsilon_B \sigma T^{3}_{eq}} \sqrt{\frac{2 \pi}{P_\mathit{rot}}}.
\end{equation}
\cite{Spencer_etal89} introduced this dimensionless parameter, which accounts for factors that affect the diurnal temperature variation. This parameter was realized by comparing the diurnal rotation of a body to the timescale in which thermal energy is stored and then re-radiated, per unit surface area \citep{Spencer_etal89}. Ignoring the effects of multiple-scattering and self-heating, this parameterization scheme changes \cref{eq3} to
\begin{equation}\label{eq15}
	\cos(i) + \Theta \frac{dT'}{dx'} \bigg|_\mathit{surf} - T'^{4}_\mathit{surf} = 0,
\end{equation}
and \cref{eq4} to
\begin{equation}\label{eq16}
	\frac{\partial T'(x',t')}{\partial t'} = \Theta \frac{\partial^{2} T'(x',t')}{\partial x'^{2}}.
\end{equation}

Surface temperatures are calculated across the surface of a sphere by solving \cref{eq16}, given the upper boundary condition \cref{eq15}. Since the amplitude of diurnal temperature changes decreases exponentially, the heat flux approaches zero, and the lower boundary becomes:
\begin{equation}
	\frac{dT'}{dx'} \bigg|_{x'\to\infty} = 0.
\end{equation}
Using the parameterized depth, time, and temperature, the number of input variables in this model is effectively reduced to two (the thermal parameter and sub-solar latitude). A finite-difference approach is used to numerically implement \cref{eq16}, as detailed in \ref{appB}. This spherical, smooth-surface TPM consisted of 13 latitude bins and was run for 46 values of sub-solar latitude ($0^{\circ}$ to $90^{\circ}$ in $2^{\circ}$ increments) and 116 values of the thermal parameter (spaced equally in logarithmic space, from 0 to 450) in order to generate the surface temperature look-up tables of $T'$.

\subsection{Rough TPM}
Our rough surface cratered TPM is similar to that originally presented by \cite{Hansen77a} which was improved upon by both \cite{Spencer90} and \cite{Emery_etal98} to include heat conduction, multiple scattering of insolation, and re-absorption of thermal radiation within spherical craters. Following the procedure of \cite{Emery_etal98}, craters are constructed with $m = 40$ planar elements contained within $k = 4$ rings that are radially symmetric about the crater center. The $k^{th}$ ring outward contains $4k$ elements, all of which are forced to have the same surface area \cite[see Figures 1 \& 2 in][for crater depiction]{Emery_etal98}. As was mentioned by \cite{Spencer90}, craters with more than 4 rings significantly increase the overall computational time and do not enhance the model resolution over the 4-ringed crater in most cases. The overall geometry of these craters is characterized by the half-opening angle, $\gamma$, as measured from the center-line of the crater to the edge (e.g. a hemispherical crater has $\gamma = 90^\circ$). The overall degree of surface roughness is characterized by the mean surface slope \citep[$\bar{\theta}$;][]{Hapke84}, which is only a function of $\gamma$ and the fraction of area covered by craters, $f_R$ \citep{Lagerros96}:
\begin{equation}
  \tan{\bar{\theta}} = \frac{2f_R}{\pi}\frac{\sin({\gamma})-\ln[1+\sin({\gamma})]+\ln\cos({\gamma})}{\cos({\gamma})-1}
\end{equation}
As shown in \cite{Emery_etal98}, the fraction of energy transferred to one crater element from another is (conveniently) equal among all elements. The scattered solar and re-absorbed thermal radiation received by the $i^{th}$ crater facet from the $j^{th}$ facet are
\begin{equation}
  E^\mathit{solar}_{i,\mathit{scat}} = \frac{S_{\odot}(1-A)}{R^{2}_{AU}} \frac{A}{1-A\frac{\gamma}{\pi}} \frac{1-\cos(\gamma)}{2m} \sum \limits^{m}_{i\neq j}\cos({i_j})
\end{equation}
and
\begin{equation}
  E^\mathit{therm}_{i,\mathit{abs}} = \frac{1-\cos(\gamma)}{2m} (1-A_{th}) \sum \limits^{m}_{i\neq j} \varepsilon_B \sigma T^{4}_{j},
\end{equation}
respectively. The energy balance at the surface (\cref{eq3}), in our parameterized time, temperature, and depth environment, becomes:
\begin{equation}
  \cos(i)(1-s) + \frac{1-\cos({\gamma})}{2m}\Bigg(\frac{A}{1-A\frac{\gamma}{\pi}} \sum \limits^{m}_{i\neq j}\cos({i_{j}}) + (1-A_{th})\sum \limits^{m}_{i\neq j} T'^{4}_{j}\Bigg) + \Theta \frac{dT'}{dx'} \bigg|_\mathit{surf} - T'^{4}_{surf} = 0.
\end{equation}
Since planetary surfaces are highly absorbing at infrared wavelengths, the Bond albedo at thermal-infrared wavelengths ($A_{th}$) is assumed to be zero. Thus, only singly-scattered re-aborption of thermal emission within the crater is considered, in contrast to multiple scattering and re-absorption of the insolation. Temperature lookup tables are generated for craters of the same values of sub-solar latitude and thermal parameter as the smooth surface TPM runs. This was done for three sets of craters of varying opening angle in order to simulate changes in roughness. Unlike the parameterized version of the smooth surface energy balance equation, $A$ is included as an independent input parameter, requiring us to explicitly account for changes in this parameter. Fortunately, it only appears in the solar scattering term, which is not a major contributor to the total energy budget. Our spherical, rough-surface TPM consists of 13 different latitude bins and was run for 3 values of $\gamma =\{ 45^{\circ}, 68^{\circ}, 90^{\circ}\}$, 46 values of sub-solar latitude ($0^{\circ}$ to $90^{\circ}$ in $2^{\circ}$ increments), 116 values of the thermal parameter (spread out in logarithmic space, from 0 to 450), and 7 values of $A^\mathit{grid} = \{0, 0.1, 0.2, 0.3, 0.4, 0.5, 1\}$\footnote{We omit values of $A$ between 0.5 and 1 since asteroid bond albedos are rarely this large.} to construct this large set of $T'$ lookup tables.

\subsection{Non-spherical Shapes}\label{subsec:32}

\begin{figure}[H]
\centering
	\includegraphics[width=10cm, clip=true]
	{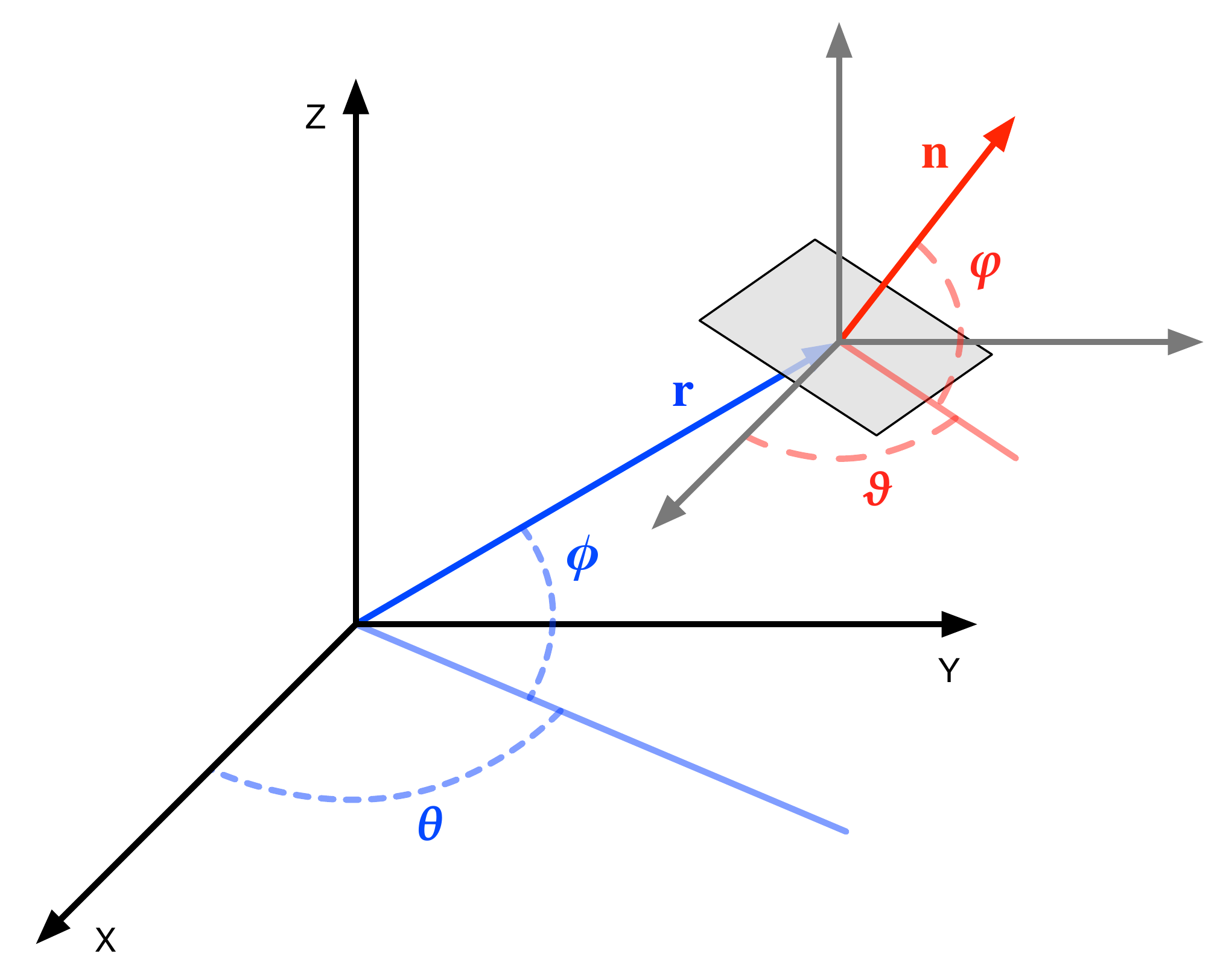}
	\caption{The coordinate systems used for a hypothetical planar facet, shown in grey. In blue: body-centric longitude and latitude, $\theta$ and $\phi$, repsectively. In red: the surface-normal longitude and latitude, $\vartheta$ and $\varphi$, respectively.}\label{fig2}
\end{figure}

Shape model facet temperatures are independent of one another in our TPM because global self-heating does not occur on the spherical and ellipsoidal shapes considered here. Since the insolation upon a facet is only dependent on the orientation of the surface normal to the sun, it is possible to map, or transform, surface temperatures from one shape to another. We exploit this fact in order to map the surface temperatures from a spherical body to convex DAMIT shape models and ellipsoids. We make use of two kinds of coordinate systems, which are depicted in \cref{fig2}. The {\it body-centric} ($\theta$, $\phi$) coordinate system defines the latitude and longitude of a facet relative to the center of the shape model. Shape model facets are often described using sets of vectors, {\bf r}, given in body-centric coordinates. Alternatively, ($\vartheta$, $\varphi$) describes a coordinate system that describes the tilt of a facet relative to the local surface, which we call the {\it surface-normal} coordinates. Both $\vartheta$ and $\varphi$ are calculated using the surface normal vector {\bf n} shown in \cref{fig2} (for a sphere, the body-centric and facet-normal coordinates are equivalent). For any given facet, on any shape model, with $\vartheta$ and $\varphi$, the temperatures can be equated to a facet on a sphere that has the same surface-normal coordinates. In \ref{appC} we derive closed-form analytic expressions to map temperatures from a sphere to the effective coordinates of a triaxial ($a \geq b \geq c$) ellipsoid.

\subsection{Flux Calculation}

Thermal flux is calculated by a summation of the individual flux contributions from smooth surface and crater elements visible to the observer (i.e. \cref{eq9}). To calculate the flux of an object not having exact $\Theta$ or sub-solar latitude values included in our lookup temperature tables, we calculate the fluxes for the closest grid points and perform a linear interpolation to compute the fluxes for the desired parameters. Instead of doing the same for $A$, we calculate the flux for a value from $A^\mathit{grid}$ and then multiply by an adjustment factor, $\Lambda$. This approach saves time and computational cost by not computing an interpolation across three variables (or dimensions). A similar approach was described by \cite{Wolters_etal11}, in which they calculated fluxes for a perfectly absorbing surface and employed a correction factor based on the desired $A$. Here, $\Lambda$ was tested empirically and adjusts the model flux based on the blackbody $T_{eq}$ curves of the desired $A$ and the closest value of $A^{grid}$.
\begin{equation}\label{eq18}
	\Lambda (A,A^{grid}) = \frac{1}{2} \Bigg[1+ \bigg(\frac{B(\lambda,T_{eq}(A^{grid}))}{B(\lambda,T_{eq}(A))}\bigg)^{2/3} \Bigg].
\end{equation}
The $\chi^{2}$ goodness-of-fit statistic is evaluated  using the modeled flux points, $F_m(\lambda)$, the observed flux measurements, $F_o(\lambda)$, with the associated 1-$\sigma$ uncertainty, $\sigma_o$, and in general:
\begin{equation}\label{eq11}
	\chi^{2} = \sum \limits \frac{(F_m(\lambda)-F_o(\lambda))^2}{\sigma_o^{2}}.
\end{equation}
We note here that instead of individual flux measurements, the modeled and observed fluxes used in \cref{eq11} represent only the extracted mean and peak-to-trough range of thermal lightcurve fluxes, and their uncertainties as calculated via error propagation. As part of the data-fitting routine, the free-parameters of the model are varied in order to minimize the goodness-of-fit as described in the following subsection.

A weighted-sum of fluxes to simulate a surface that is comprised of smooth and rough surface patches:
\begin{equation}\label{eq9}
	F(\lambda) = \frac{\varepsilon_{\lambda}}{\Delta_{AU}^{2}} \int\limits_{\ S} \!\!\! \int \{(1-f_R)B_{smooth}(\lambda,T(\theta,\phi)) + f_R(1-v)\Lambda B_{rough}(\lambda,T(\theta,\phi,i))\} \cos(e) dA.
\end{equation}
in which $\Delta_{AU}$ is the observer-centric distance in AU. The visibility factor, $v$, is analogous to the shadowing factor; $v=1$ if a facet is hidden from view, otherwise $v=0$. Each point on the surface is treated as a blackbody emitter with wavelength-dependent emissivity, $\varepsilon_{\lambda}$ and temperature, $T = T_{eq}T'$:
\begin{equation}\label{eq10}
	B(\lambda,T) = \frac{2hc^{2}}{\lambda^{5}}\frac{1}{\exp(hc/\lambda k_{b}T)-1}.
\end{equation}

\subsection{Data-Fitting Routine}\label{subsec:33}

At each epoch, both the mean flux and peak-to-trough range of the thermal light curve are extracted and used in the model fitting procedure. As alluded to in \cref{sec2}, these two parameters contain diagnostic information about the object's shape, spin direction, and thermophysical properties and can be easily extracted from non-dense thermal lightcurve data. In principle, it is just as feasible to fit models to each independent flux point that contributes to an object's infrared lightcurve. Doing so could offer insight into the shape, as departures from a sinusoidal lightcurve can indicate relative topographic lows or highs. However, such an approach works best in cases in which the thermal lightcurve is densely sampled, which is often not the case in untargeted astronomical surveys such as IRAS, Akari, and WISE. In this work we focus on estimating only the elongation of a body by incorporating the photometric range (peak-to-trough) of the thermal lightcurve. TPM fitting to sparse thermal lightcurves is prone to systematic parameter bias from oversampling and/or heteroscedastic uncertainties across many rotational phases, which can unevenly emphasize certain rotational phases over others, thus skewing the best-fit parameters. An example of this would be a scenario in which peaks of the modeled lightcurce were fit better with the observed lightcurve minima, thus resulting in an overall lower best-fit modeled fluxes. However, the capacity and capability of incorporating thermal lightcurve data into the shape model inversion process, when it is combined with visible lightcurve data, should be explored further \citep{Durech_etal14,Durech_etal15}.

In our data fitting approach, the shape, spin vector ($\lambda^{eclip}_{l}$, $\beta^{eclip}_{l}$), roughness, and thermal inertia are left as free parameters that we select from a pre-defined sample space, and we search for the best-fit $D_\mathit{eff}$. A sphere and ellipsoids with b/c=1 (i.e. prolate) with a/b axis ratios of 1.25, 1.75, 2.5, and 3.5 are used. For each of these shapes, we sample through 25 predefined thermal inertia values, 3 default roughness ($\bar{\theta}$) values, and 235 spin vectors. Following \cite[table 1 in ][]{Delbo&Tanga09}, each value of $\gamma$ is paired with a corresponding $f_R = \{0.5, 0.8, 1.0\}$ value in order to produce default mean surface slopes of $\bar{\theta} = \{10^{\circ}, 29^{\circ},$ and $58^{\circ}\}$. The thermal inertia points are evenly spread in log space from 0 to 3000 Jm$^{-2}$K$^{-1}$s$^{-1/2}$, and the spin vectors are spread evenly throughout the celestial sphere. For each shape/spin vector/$\Gamma$ combination we use a routine to find the $D_\mathit{eff}$ value which minimizes $\chi^{2}$.

The grid of spin vectors are formed by constructing a Fibonacci lattice in spherical coordinates \citep{Swinbank&Purser06}. Here, a Fermat spiral is traced along the surface of the celestial sphere, using the golden ratio ($\Phi \approx 1.618$) to determine the turn angle between consecutive points along the spiral. The result is a set of points with near-perfect homogeneous areal coverage across the celestial sphere. This Fibonacci lattice constructed here uses $N=235$ points with the $l^{th}$ point ($i \in [0,N-1]$) having an ecliptic latitude and longitude of
\begin{equation}\label{eqbeta}
  \beta^\mathit{eclip}_{l} = \arcsin\bigg(\frac{2l-N+1}{N-1}\bigg)
\end{equation}
and
\begin{equation}\label{eqlambda}
  \lambda^\mathit{eclip}_{l} = 2\pi l\Phi^{-1} \bmod 2\pi,
\end{equation}
respectively. The mean flux value, and the peak-to-trough range, at each wavelength and at each epoch are taken as input to \cref{eq11}. The Van Wijngaarden-Dekker-Brent minimization algorithm, commonly known as {\it Brent's Method} \citep{Brent73}, combined with the golden section search routine \citep[\S 10.2][]{Press_etal07}, is used since it does not require any derivatives of the $\chi^2$ function to be known. The algorithm uses three values of $\chi^2$ evaluated at different input diameters to uniquely define a parabola. The algorithm used the minima of these parabolas to iteratively converge on the $D_\mathit{eff}$/$A$ combination that corresponds to the global minimum of $\chi^2$ value to within 0.01\%.

The values of $D_\mathit{eff}$ and $A$ are linked through \cref{eq12} and \cref{eq13} for a smooth surface, but the effect of multiple scattering alters the energy balance, and thus effective albedo \citep{Mueller07}\footnote{The crater opening angle used in \cite{Mueller07} is twice that used here.}, for a cratered surface:
\begin{equation}\label{eq21}
  A^\mathit{crater}(\gamma) = A\frac{1-\sin^2(\gamma/2)}{1-A\sin^2(\gamma/2)}.
\end{equation}
An object having an areal mixture of both smooth and rough topography has an {\it effective} bond albedo, $A_\mathit{eff}$, which is a weighted average of the albedo of a smooth surface and the bond albedo of a crater $A^{crater}$ \citep{Wolters_etal11}:
\begin{equation}\label{eq17}
  A_\mathit{eff}(f_R,\gamma) = (1-f_R)A + f_RA^\mathit{crater}.
\end{equation}

To place confidence limits on each of the fitted parameters, we use the $\chi^2$ values calculated during the fitting procedure. In general, the $\chi^2$ distribution depends on $\nu$ degrees of freedom (3 in the present work), which is equal to the number of data points (constraints; 8 in this work, as described in \cref{subsec:41}) minus the number of input (free; 5 in this work) parameters. Since the $\chi^2$ distribution has an expectation value of $\nu$ and a standard deviation of $\sqrt{2\nu}$, the best-fit solutions cluster around $\chi^2_\mathit{min} = \nu$, and those with $\chi^2 < (\nu + \sqrt{2\nu})$ represent 1-$\sigma$ confidence estimates\footnote{Consequently, $\chi^2 < (\nu + 2\sqrt{2\nu})$ and $\chi^2 < (\nu + 3\sqrt{2\nu})$ give the respective 2- and 3-$\sigma$ confidence limits.}. However, because of the TPM assumptions (e.g. no global self-heating, homogeneous thermophysical properties) it is often possible for the TPM to not perfectly agree with the data, in which case $\chi^2_\mathit{min} > \nu$. In this case, we use $\sfrac{\chi^2}{\chi^2_\mathit{min}} < \nu + \sqrt{2\nu}$ to place 1-$\sigma$ confidence limits. This modification effectively scales the cutoff bounds by a factor of $\chi^2_\mathit{min}$, instead of adopting the traditional approach of using a constant $\chi^2$ distance cutoff. This scaling of the $\chi^2$ cutoff bounds is a more conservative approach in quoting parameter uncertainties, as it includes the systematic uncertainties that lead to the larger $\chi^2_\mathit{min}$ in the reported parameter uncertainties. Using the {\it reduced} $\chi^2$ statistic ($\tilde{\chi}^2 = \sfrac{\chi^2}{\nu}$), we can express the solutions within a 1-$\sigma$ range as $\tilde{\chi}^2 < \tilde{\chi}^2_\mathit{min}(1+\sfrac{\sqrt{2\nu}}{\nu})$.

In \cref{sec5}, we report TPM results for all of the objects that we analyzed, even those for which a high $\chi^2_\mathit{min}$ (indicating a poor fit to the data), which are considered unreliable and should only be used with caution.


\section{Method Testing and Validation}\label{sec4}

\Cref{sec2} describes how pre- and post-opposition multi-wavelength thermal observations are able to simultaneously constrain multiple thermophysical (albedo, thermal inertia, and surface roughness) and physical (diameter, shape and spin direction) properties of an object. In this section we present results from a proof-of-concept test of the ability to constrain the above parameters with WISE  pre- and post-opposition observations. In performing this test, we generate an artificial flux dataset from shape models from DAMIT \citep[Database of Asteroid Models from Inversion Techniques;][]{Durech_etal10} as a benchmark for testing the accuracy and precision of our TPM approach. Specifically, we search for and quantify any biases that may exist among each fit parameter. We also include a comparison to the uncertainty estimates to typical values found in previous works.

\subsection{Synthetic Flux Dataset}\label{subsec:41}

\Cref{table1} lists the objects, DAMIT shape models used, the associated rotation periods and the effective diameter, as computed beforehand via a NEATM fit to the real WISE observations. The observing geometries used in the synthetic model runs are the same as the WISE observations listed in \cref{table2}. We calculate the artificial thermal emission for WISE photometric filters W3 and W4 (12 \& 24 $\mu$m) for a full rotation of the shape. From these infrared lightcurves, we extract the mean and peak-to-trough flux range for each filter as input into our fitting routine (\cref{eq11}). Since  we now have these 2 quantities, observed twice for two wavelengths, there are 8 data points to constrain 5 free parameters: diameter, thermal inertia, surface roughness, shape elongation, and sense of spin. Fluxes for each shape model are calculated using the TPM described here by using the actual WISE observing circumstances detailed in \cref{table2}. If an object has two shape models available - a result of ambiguous solutions of the lightcurve inversion algorithm - then both were included in the analysis.

\doublespacing
\begin{footnotesize}
\begin{longtable}{lccccccc}
\caption{Shape Models \& Physical Properties for Synthetic Dataset}\label{table1}\\

Object & Shape Model & a/b$^\textrm{synth}$ & Spin Vector & $P_\mathit{rot}$ (hr) & $D^\textrm{synth}_\mathit{eff}$ (km) & $H_V$ & $G_V$ \\ \hline \multicolumn{7}{c}{ } \\[-1,6em] \hline \endfirsthead

\multicolumn{8}{c}{\tablename\ \thetable{} -- continued}\\
\vspace{-0.25cm}\\ 
Object & Shape Model & a/b$^\textrm{synth}$ & Spin Vector & $P_\mathit{rot}$ (hr) & $D^\textrm{synth}_\mathit{eff}$ (km) & $H_V$ & $G_V$ \\ \hline \multicolumn{7}{c}{ } \\[-1,6em] \hline \endhead

\vspace{-0.25cm} & M171 & 1.31 & $-69^{\circ}$, $107^{\circ}$ \\
\vspace{-0.25cm}(167) Urda & & & & 13.06133 & 43.00 & 9.131 & 0.283 \\
 & M172 & 1.30 & $-68^{\circ}$, $249^{\circ}$ \\
\hline
(183) Istria & M669 & 1.39 & $20^{\circ}$, $8^{\circ}$ & 11.76897 & 35.48 & 9.481 & 0.221 \\
\hline
\vspace{-0.25cm} & M182 & 1.22 & $-75^{\circ}$, $20^{\circ}$ \\
\vspace{-0.25cm}(208) Lacrimosa & & & & 14.0769 & 44.35 & 9.076 & 0.232 \\
 & M183 & 1.23 & $-68^{\circ}$, $176^{\circ}$ \\
\hline
(413) Edburga & M354 & 1.37 & $-45^{\circ}$, $202^{\circ}$ & 15.77149 & 35.69 & 9.925 & 0.296 \\
\hline
\vspace{-0.25cm} & M521 & 1.41 & $24^{\circ}$, $90^{\circ}$ \\
\vspace{-0.25cm}(509) Iolanda & & & & 12.29088 & 60.07 & 8.476 & 0.382 \\
 & M522 & 1.32 & $54^{\circ}$, $248^{\circ}$ \\
\hline
(771) Libera & M250 & 1.50 & $-78^{\circ}$, $64^{\circ}$ & 5.890423 & 29.52 & 10.28 & 0.323 \\
\hline
\vspace{-0.25cm} & M609 & 2.30 & $34^{\circ}$, $38^{\circ}$ \\
\vspace{-0.25cm}(857) Glasenappia & & & & 8.20756 & 15.63 & 11.29 & 0.246 \\
 & M610 & 2.32 & $48^{\circ}$, $227^{\circ}$ \\
\hline
(984) Gretia & M256 & 1.57 & $52^{\circ}$, $245^{\circ}$ & 5.778025 & 35.76 & 9.526 & 0.379 \\
\hline
(1036) Ganymed & M261 & 1.05 & $-78^{\circ}$, $190^{\circ}$ & 10.313 & 37.42 & 9.236 & 0.311 \\
\hline
\vspace{-0.25cm} & M403 & 1.76 & $-73^{\circ}$, $12^{\circ}$ \\
\vspace{-0.25cm}(1140) Crimea & & & & 9.78693 & 31.77 & 9.621 & 0.207 \\
 & M404 & 1.61 & $-22^{\circ}$, $175^{\circ}$ \\
\hline
(1188) Gothlandia & M479 & 1.71 & $-84^{\circ}$, $334^{\circ}$ & 3.491820 & 13.29 & 11.52 & 0.254 \\
\hline
\vspace{-0.25cm} & M409 & 2.03 & $35^{\circ}$, $106^{\circ}$ \\
\vspace{-0.25cm}(1291) Phryne & & & & 5.584137 & 27.87 & 10.29 & 0.251 \\
 & M410 & 2.30 & $59^{\circ}$, $277^{\circ}$ \\
\hline
\vspace{-0.25cm} & M657 & 1.27 & $54^{\circ}$, $225^{\circ}$ \\
\vspace{-0.25cm}(1432) Ethiopia & & & & 9.84425 & 7.510 & 12.02 & 0.282 \\
 & M658 & 1.32 & $44^{\circ}$, $41^{\circ}$ \\
\hline
(1495) Helsinki & M656 & 1.74 & $-39^{\circ}$, $355^{\circ}$ & 5.33131 & 13.54 & 11.41 & 0.359 \\
\hline
(1568) Aisleen & M422 & 2.30 & $-68^{\circ}$, $109^{\circ}$ & 6.67597 & 13.60 & 11.49 & 0.131 \\
\hline
\vspace{-0.25cm} & M477 & 1.56 & $70^{\circ}$, $222^{\circ}$ \\
\vspace{-0.25cm}(1607) Mavis &&  & & 6.14775 & 15.10 & 11.32 & 0.256 \\
 & M478 & 1.93 & $59^{\circ}$, $0^{\circ}$ \\
\hline
(1980) Tezcatlipoca & M274 & 1.72 & $-69^{\circ}$, $324^{\circ}$ & 7.25226 & 5.333 & 13.57 & 0.186 \\
\hline
(2156) Kate & M438 & 2.15 & $74^{\circ}$, $49^{\circ}$ & 5.622153 & 8.678 & 4.339 & 0.186 \\
\hline
\vspace{-0.25cm} & M704 & 1.68 & $-50^{\circ}$, $197^{\circ}$ \\
\vspace{-0.25cm}(4611) Vulkaneifel & & & & 3.756356 & 12.38 & 11.87 & 0.268 \\
 & M705 & 1.64 & $-86^{\circ}$, $5^{\circ}$ \\
\hline
\vspace{-0.25cm} & M682 & 3.48 & $-78^{\circ}$, $97^{\circ}$ \\
\vspace{-0.25cm}(5625) 1991 AO$_2$ & & & & 6.67412 &7 14.25 & 12.83 & 0.165 \\
 & M683 & 3.18 & $-52^{\circ}$, $265^{\circ}$ \\
\hline
\vspace{-0.25cm} & M757 & 1.55 & $67^{\circ}$, $62^{\circ}$ \\
\vspace{-0.25cm}(6159) 1991 YH & & & & 10.65893 & 5.148 & 13.38 & 0.175 \\
 & M758 & 1.65 & $67^{\circ}$, $266^{\circ}$ \\
\hline

\end{longtable}
\end{footnotesize}
\singlespacing

\subsection{Model Validation Results}\label{subsec:42}

\begin{figure}[th!]
  \centering
  \begin{tabular}[b]{rl}
	\includegraphics[width=.5\linewidth]{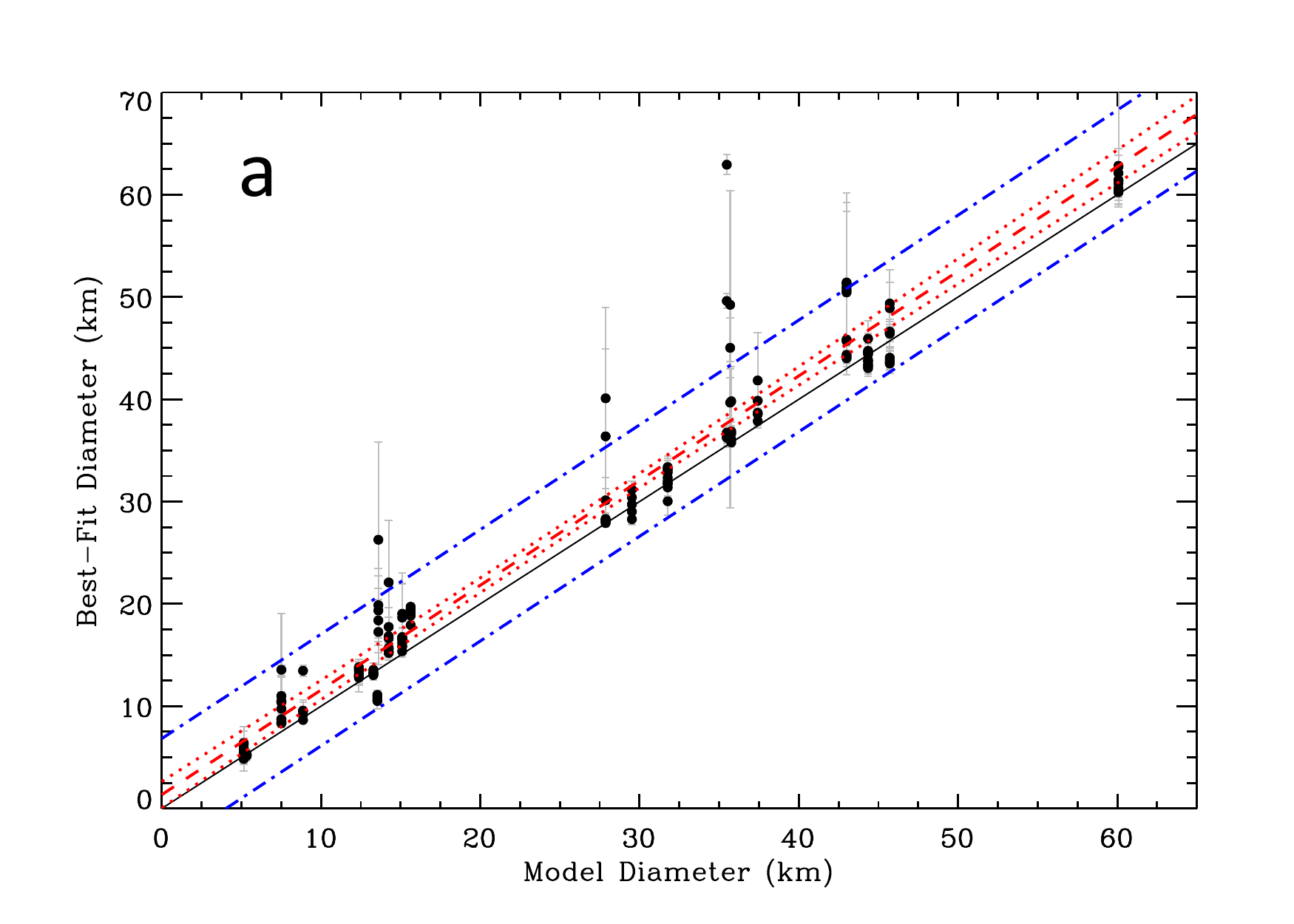} &
	\includegraphics[width=.5\linewidth]{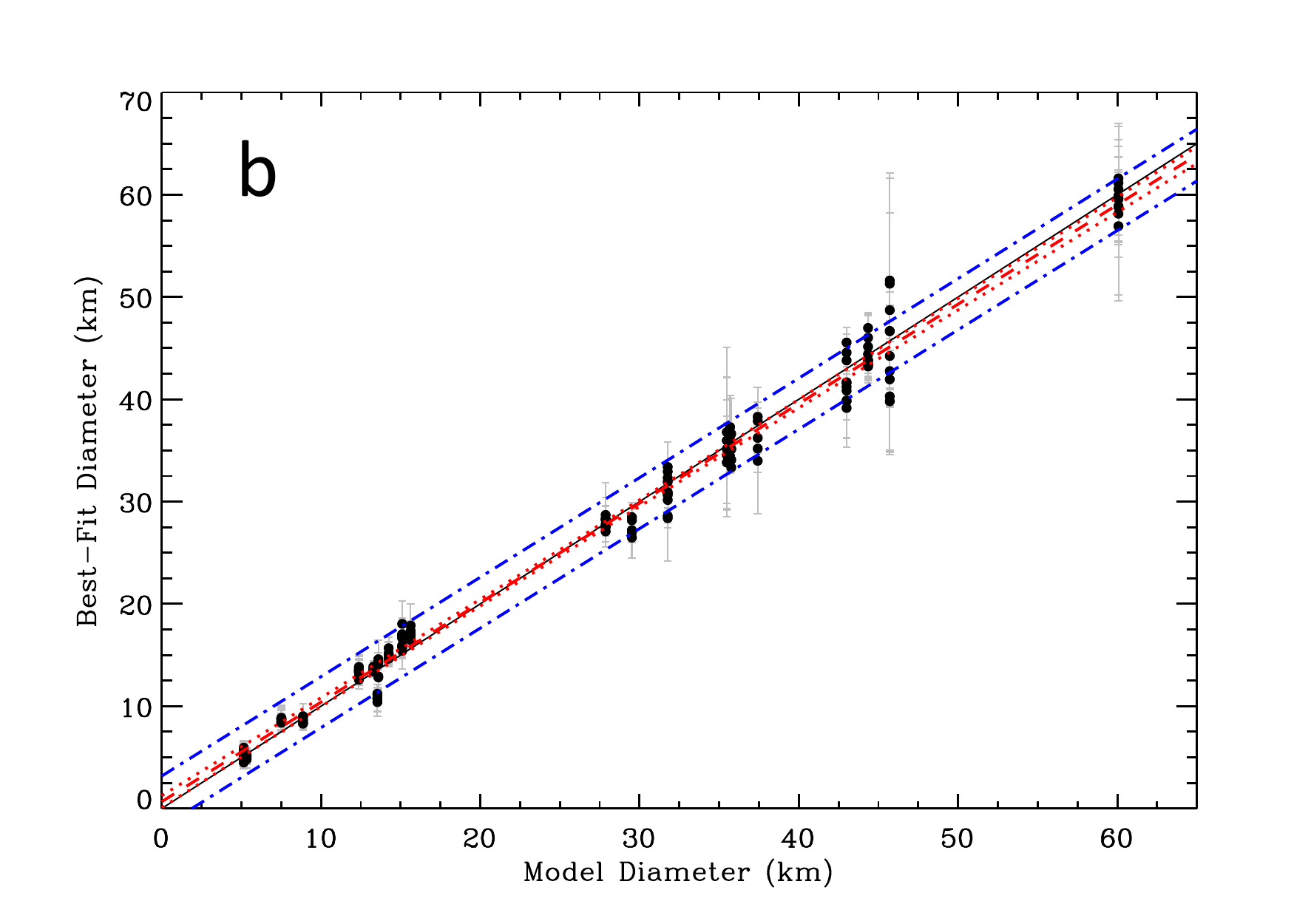} \\
	\includegraphics[width=.5\linewidth]{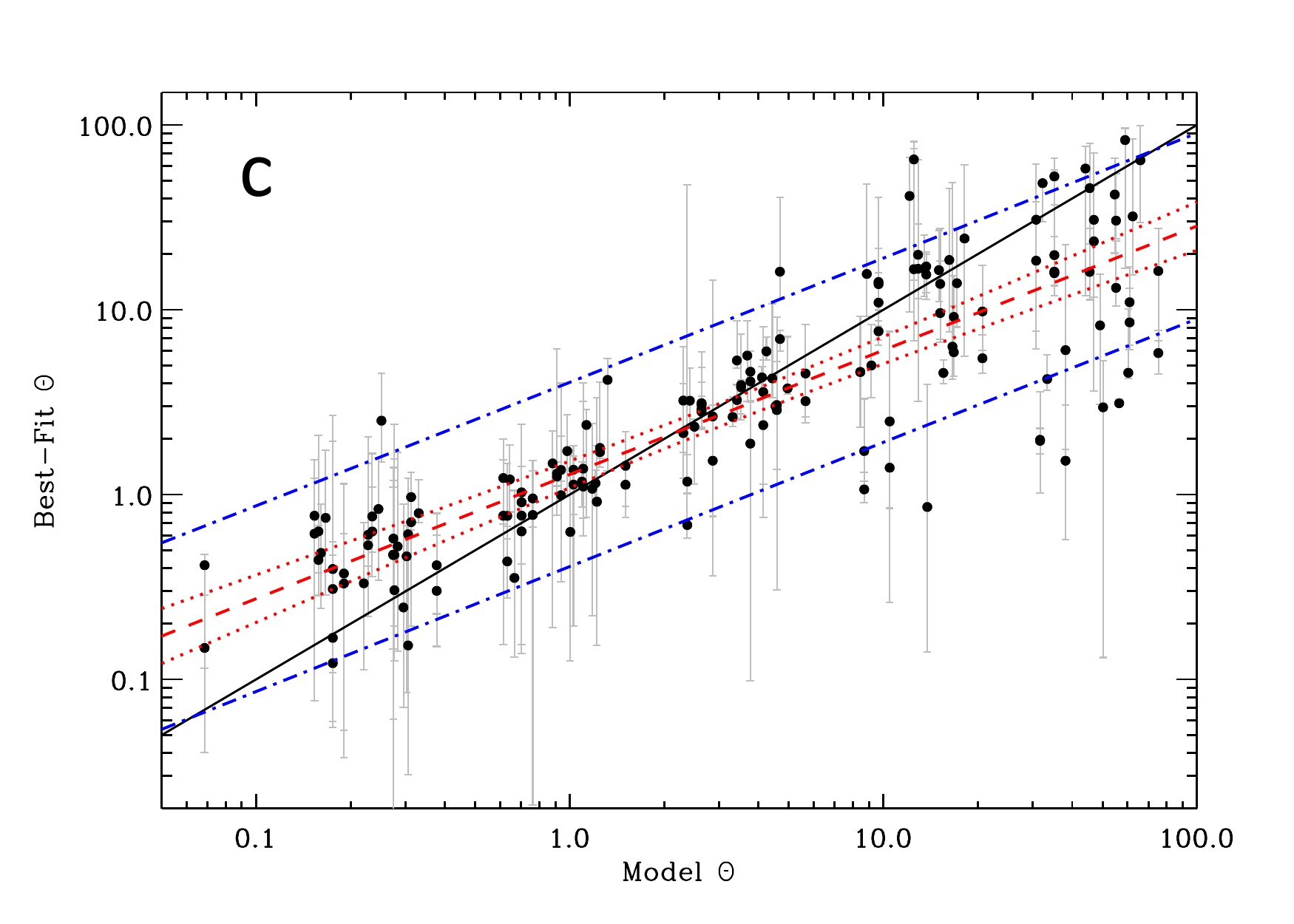} &
	\includegraphics[width=.5\linewidth]{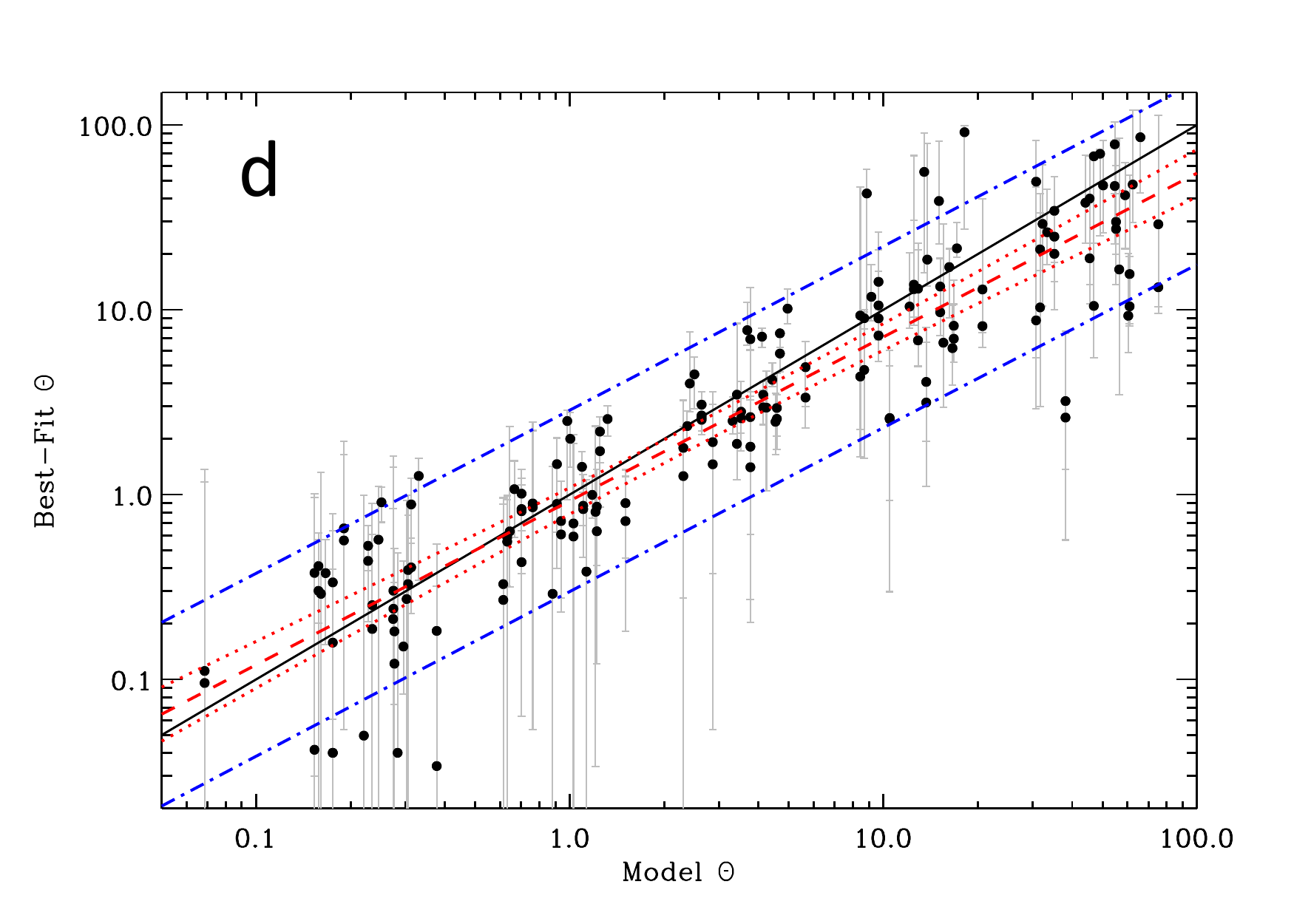} \\
	\includegraphics[width=.5\linewidth]{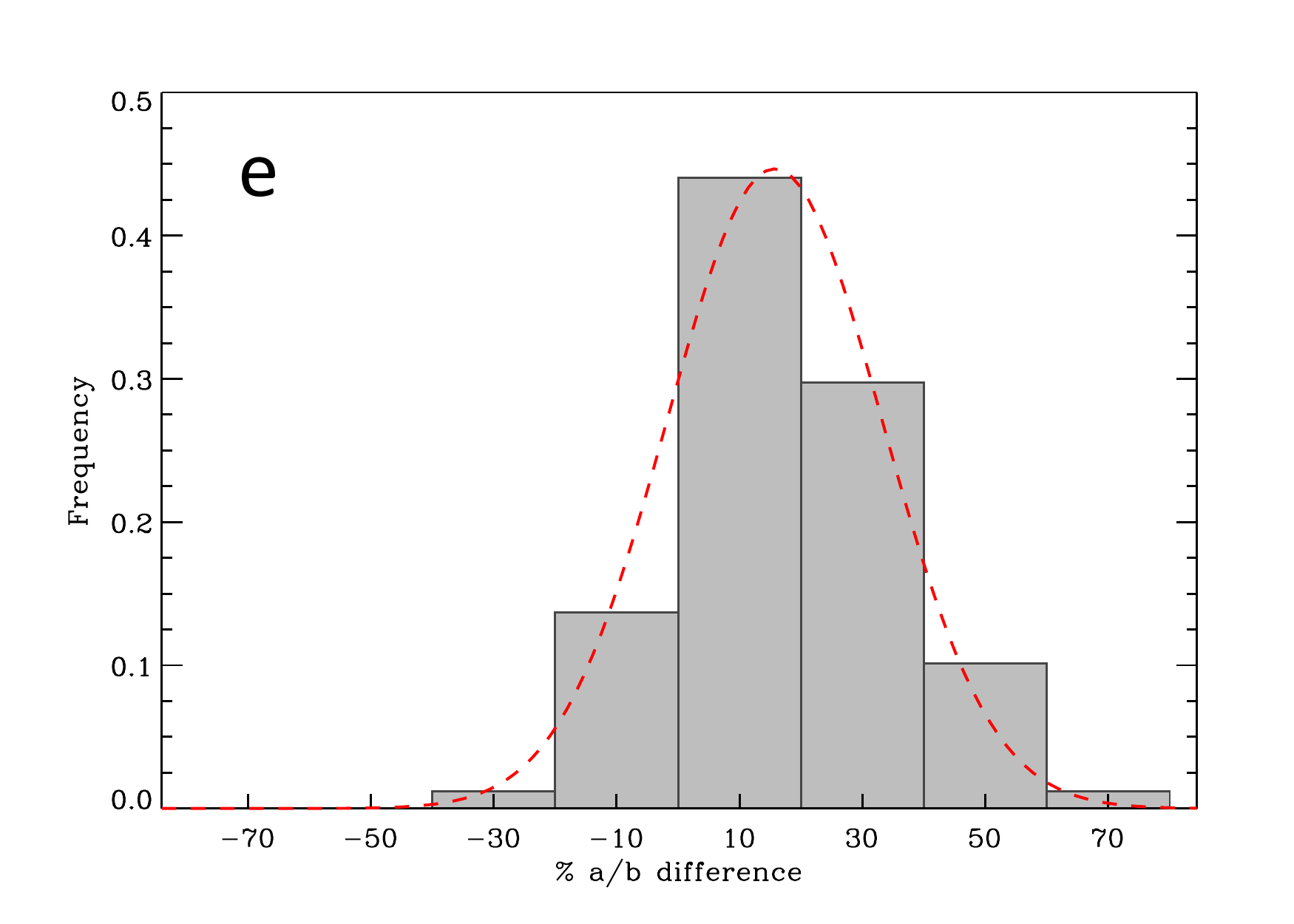} &
	\includegraphics[width=.5\linewidth]{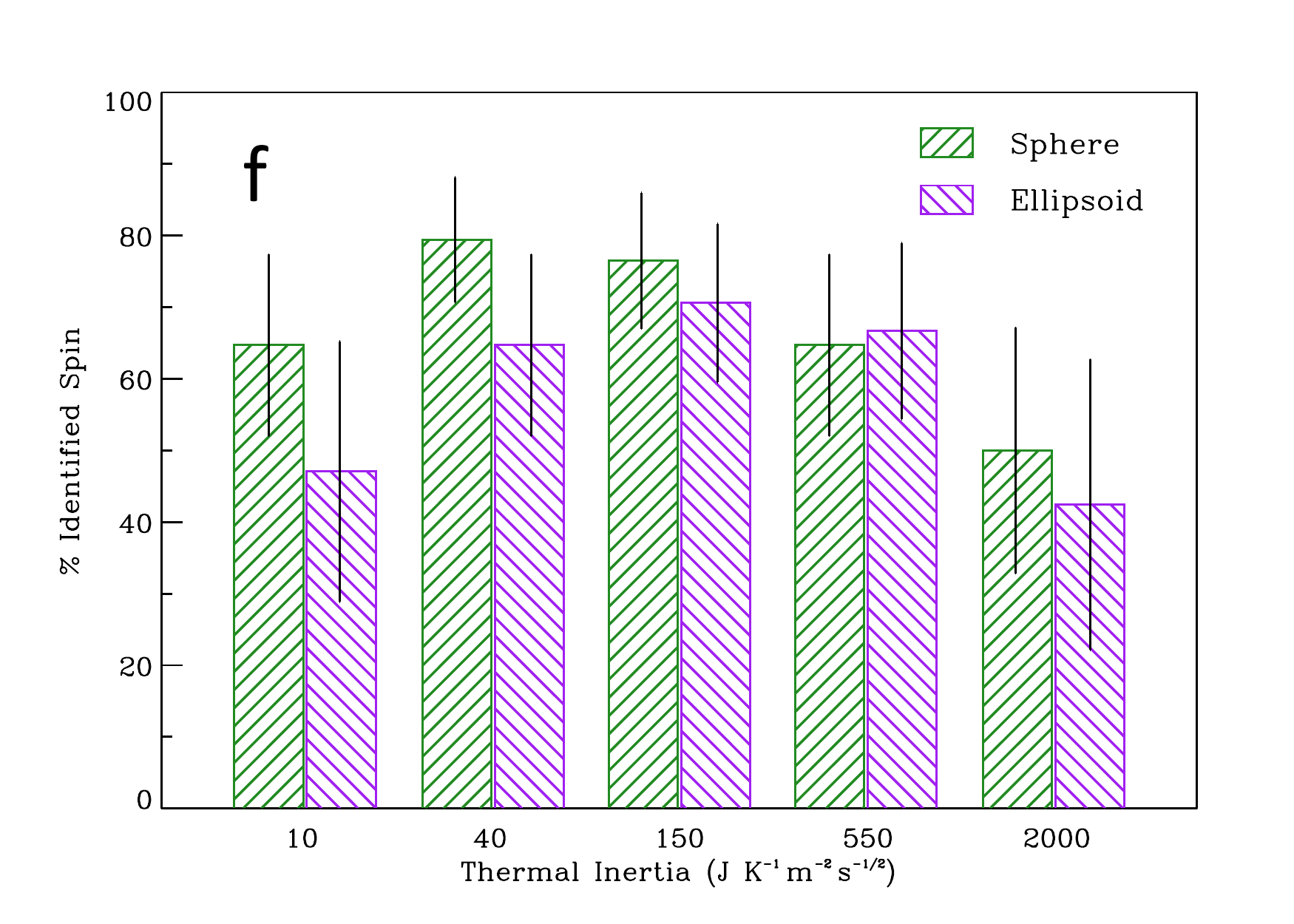} \\
  \end{tabular}
  \caption{The best-fit diameter vs. model diameter for spherical (a) and ellipsoidal (b) TPM shape. Panels (c) and (d) show the same for $\Theta$. The red dashed and dotted lines show the best fit to the data and the corresponding 95\% confidence interval. Blue, dash-dotted lines show the 68\% prediction interval. The best-fit thermal parameter vs. model thermal parameter. Panel (e) shows the fractional difference in DAMIT area-equivalent a/b to the best fit ellipsoid, expressed as a percentage. Lastly, the percentage of TPM solutions that correctly identifiy the spin direction is given in panel (f).}\label{fig3}
\end{figure}

The best-fit TPM parameter results for the synthetic dataset are detailed below and depicted in \cref{fig3}. Overall, using ellipsoid shapes results in more accurate and precise estimates for diameter, thermal inertia, and shape.

\paragraph{Diameter Constraints} \Cref{fig3} includes diameter uncertainties, shown by grey bars, based off of the noise of WISE data. The uncertainty overlap with the expected diameter values indicates that the uncertainty in the data is equal to or larger than the uncertainty introduced by model assumptions. The assumption of a spherical shape results in overestimation of diameter of up to 20$\%$, as shown in \cref{fig3}(a). The TPM performs better when using ellipsoid shapes; at most, diameters are overestimated by 10$\%$. These offsets are particularly pronounced for highly elongated objects with high $\Theta$ values and when observed at large sub-solar latitude values. In these cases, the average observed cross-sectional area is particularly large and surface temperatures are, on average, warmer (\cref{fig1}(c)), since a larger fraction of the surface experiences perpetual daylight. Diameter uncertainties when using ellipsoid shapes are consistent with the $\pm10\%$ value seen in other thermophysical modeling papers, yet seem to imply a drop in accuracy from the NEATM \citep{Harris06}. However, care must be taken when comparing the performance results presented here, which are based on synthetic data generated from shape models with various spin vectors, and those presented by \cite{Harris06}, which uses an idealized synthetic dataset generated from spherical shapes with no obliquity. The work of \cite{Wright07}, who included rough spheres in their performance test of NEATM, shows diameter accuracy of the NEATM to be on par with TPM works, such as this one.

\paragraph{Thermal Inertia/Parameter Constraints} We transform thermal inertia estimates into thermal parameter space \cref{eq14}; since the rotation periods of synthetic objects affect the temperature distribution and must be accounted for. Panels (c) and (d) in \cref{fig3} show the estimated $\Theta$ values for the spherical and best-fit ellipsoidal shape, respectively, against the synthetic values. When a sphere is used for a TPM, thermal inertia is overestimated at low values and underestimates at large $\Theta$ values. At either extreme of $\Theta$, this offset is systematically different by a factor of $\sim$4. Ellipsoidal shapes do a much better job matching the input $\Theta$, with systematic offsets of less than 25\% (red dashed line in \cref{fig3}(d)). We investigate this discrepancy further in \cref{subsec:43} and note here that the estimated uncertainties (grey bars), calculated from WISE signal-to-noise ratios for each object, are far larger than the systematic offset, indicating that  the assumptions in our TPM fitting do not contribute much to the overall uncertainty in thermal inertia and the $\chi^2$-based errors values reasonably reflects the overall precision in our TPM.

\paragraph{Shape Constraints} To assess how well our method is able to constrain the shape (elongation) of an object, we compare the best-fit ellipsoid a/b to that of the area-equivalent a/b ratio of the DAMIT shape model. Since we use a sparse grid of possible a/b values, placing meaningful confidence bars based on $\chi^2$ values is not practical. Instead, we plot the frequency distribution of $\%$a/b (= $\frac{a/b_\textrm{ellip}}{a/b_\textrm{DAMIT}}$ - 1) difference in \cref{fig3}. Fitting a Gaussian function (16\%, with standard deviation of 18\%) to the distribution shows that the assumptions within the model, most likely the use of a prolate ellipsoid (a/c = 1), results in a slight overestimation of the a/b axis. The standard deviation of this distribution is reflective not of the uncertainty in the mean offset in the shape accuracy, but reflective of the inherent model uncertainties that arise from assuming an ellipsoid shape. Accounting for this discrepancy, we shift the TPM best-fit shape result downward by 16\% and assign an uncertainty of 18\%.

\paragraph{Spin Constraints} We also investigated the ability to constrain the spin direction of an object (i.e. retrograde or prograde rotation). In total, the ellipsoid TPM correctly identified the spin direction $58.3 \pm 6.5$\% of the time compared to $67.1 \pm 5.7$\% for the spherical TPM. In \cref{subsec:25} we pointed out that thermal inertia is a significant factor when constraining spin direction. Thus, we break down our results into the thermal inertia bins assigned for the artificial dataset. \Cref{fig3}(f) shows the percentage of best-fit solutions, broken down by sphere/ellipsoid shape assumption, in which spin direction was correctly identified for a given thermal inertia. It is clear that intermediate values of thermal inertia provide more reliable constraints on the spin direction, which can be explained by the increased asymmetry of thermal phase curves at intermediate values of thermal inertia. As more extreme (low and high) values of thermal inertia result in more symmetric thermal phase curves, which make it difficult to distinguish morning and afternoon hemispheres, and thus spin direction (\cref{fig1}(a)).

\doublespacing
\begingroup
\begin{longtable}{r|cc|cc}
\caption{Multiple Linear Regression Results}\label{table5}\\
	& \multicolumn{2}{c}{\%$\Delta D_\mathit{eff}$}  & \multicolumn{2}{c}{\%$\Delta\Theta$} \\
  & coefficient & p-value & coefficient & $p$-value \\
 \hline
 D$^\textrm{synth}_\mathit{eff}$ & \cellcolor{Gray} -0.0009 $\pm$ 0.0004 & \cellcolor{Gray} 0.02 & 0.0012 $\pm$ 0.0017 & 0.48 \\ 
 log$_{10}(\Theta^\textrm{synth})$ & -0.0037 $\pm$ 0.0067 & 0.59 & \cellcolor{Gray} -0.1390 $\pm$ 0.0292 & \cellcolor{Gray} $< 0.01$ \\ 
 s-s lat.$^\textrm{synth}$ & | & | & -0.0036 $\pm$ 0.0024  & 0.12 \\
 s-o lat.$^\textrm{synth}$ & \cellcolor{Gray} 0.0024 $\pm$ 0.0005 & \cellcolor{Gray} $< 0.01$ & | & | \\
 a/b$^\textrm{synth}$ & 0.0085 $\pm$ 0.0119 &0.48 & 0.0573 $\pm$ 0.0507 & 0.26 \\
 intercept & -0.0264 $\pm$ 0.0319 &0.41 & -0.0621 $\pm$ 0.1357 & 0.65 \\
 
\end{longtable}
\endgroup
\singlespacing

\subsection{Parameter Bias Analysis}\label{subsec:43}

We performed a linear multiple regression analysis on the percent diameter and $\Theta$ difference ($\%\Delta D_\mathit{eff}$ and $\%\Delta\Theta_\mathit{eff}$, between the synthetic and TPM estimates) to distinguish, if any, factors which bias the estimates. The predictor variables chosen were log$_{10}(\Theta)$, the sub-solar latitude (s-s lat.) or sub-observer latitude (s-o lat.), and the elongation of the shape model (a/b$^\textrm{synth}$), because they are the most likely to affect the response variable (\cref{sec2}). The regression models the response variable - the percent difference in diameter or $\Theta$ - as being linearly dependent on the sum of any number of independent predictor variables - in this case 4. For each predictor, the regression model computes a slope coefficient which represents the change in that variable when all others are held constant. An intercept term, quantifying the value of the response when all predictors are zero, is also computed. \Cref{table5} shows the slope coefficients, intercepts, along with the associated p-values; indicating the statistical significance of each predictor variable in the multiple regression model. We use $p < 0.05$, representing a 95\% confidence level, to identify predictor variables that affect the response variable (p-values that are larger than this cutoff indicate that the predictor variable has little to no statistically relevant effect on the response variable).

\begin{figure}[H]
\centering
	\includegraphics[width=10cm, clip=true]
	{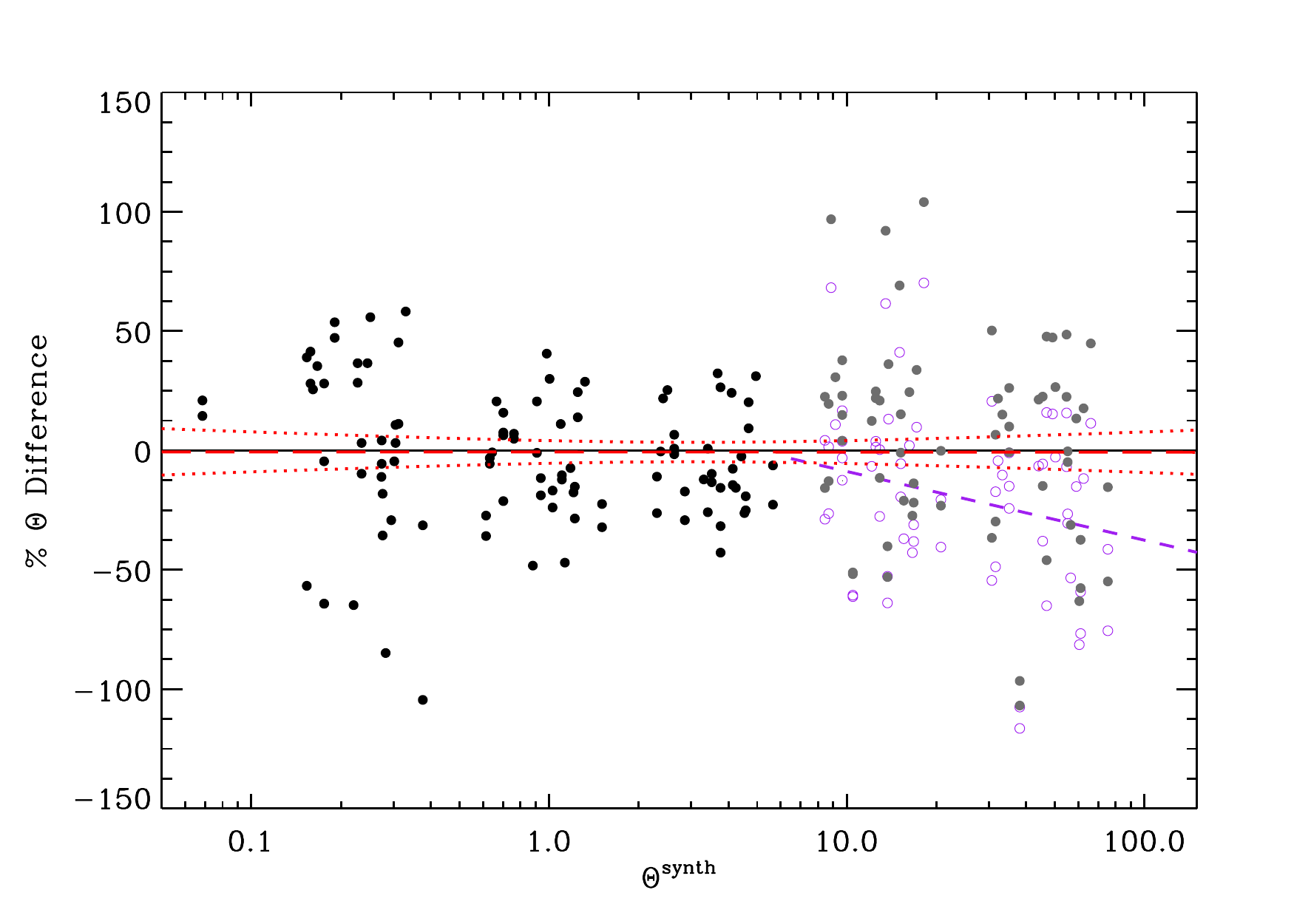}
	\caption{Percent difference between synthetic and best-fit $\Theta$ as a function of the synthetic (known) value. For $\Theta >$ 6, we derive a formula (\cref{eq24}) to adjust the underestimation (open purple circles) such that the corrected values (grey dots) are symmetric about zero percent difference, represented by the black horizontal line. The red best-fit line to the black and grey dots has a slope of nearly zero and shows no evidence for a bias at either $\Theta$ extreme.}\label{fig4}
\end{figure}

The diameters calculated from our approach are overestimated when observations occur at high sub-observer latitudes ($p < 0.01$) and for small diameters ($p = 0.02$) as marked in grey in \cref{table5}. From the multiple regression results performed on $\Theta$, the only predictor that explains the variance in response variable is $\Theta^\textrm{synth}$, marked in \cref{table5} with grey. The negative sign represents an overestimation for small $\Theta$ and/or underestimation at large $\Theta$. A re-examination of panel (d) in \cref{fig3} shows that this offset is due to points with $\Theta >$ 6. For lower values of $\Theta,$ the best-fit line is likely skewed upward due to the underestimation at the larger end. We thus, employ a formulation to correct objects for which $\Theta >$ 6:
\begin{equation}\label{eq24}
	\Theta^\mathit{corr} = 10^{({\frac{b}{1+m}})} \Theta^{(\frac{1}{1+m})}.
\end{equation}
To determine correction factor, we fit a line to objects with $\Theta > 4$ (shown by the purple-dotted line in \cref{fig4}) and compute $m = -0.136$ and $b = 0.026$. This fit provides a means of removing the systematic underestimation in thermal parameter, and by proxy, thermal inertia. The bias seen in panel (d) of \cref{fig3} vanishes, as shown by the red best-line in \cref{fig4} re-computed after using \cref{eq24} to adjust the $\Theta^\textrm{synth} >$ 6 values. Thermal parameters this large are more likely to be found for objects with unusually high thermal inertia and/or among slow rotators - or for icy bodies low surface temperatures in the outer solar system \citep[e.g. table 1 in][]{Spencer_etal89}.

\subsection{Other Possible Sources of Bias}

Upon closer inspection of the multiple regression results we see that the skew at small sizes is due to particular objects being observed at high latitudes rather than by the intrinsic diameter of the object. The effect that the viewing geometry has on the size estimation can be explained by the fact that the thermal lightcurve-averaged flux does not properly represent the effective diameter. For the extreme case of observations taken from a ``pole-on'' geometry, the thermal flux represents the largest feasible cross-sectional area, which would result in the overestimation of diameters, regardless of the shape used in the TPM.

In some of the largest publications of independently-derived thermal inertias \citep{Delbo&Tanga09,Hanus_etal15,Hanus_etal18}, convex shape models were used in the TPM approach. One critique of using convex shape models is that actual asteroids can potentially harbor shape concavities, which raises potential concern as to whether or not large deviations from a spherical shape will bias the temperature distribution in any significant way \citep{Rozitis&Green13}. Radar observations of the NEA (341843) 2008 $\textrm{EV}_5$ showed a large concavitiy, and thermal observations were analyzed in detail by \cite{AliLagoa_etal13}. They found no evidence that the concavity influenced the results, as its effects on the thermal emission (i.e. shadowing and global self-heating) was likely below the signal-to-noise of the observations. Using a larger set of shape models with concavities \cite{Rozitis&Green13} demonstrated that the effects of global self-heating and shadowing have negligible effects on the temperature distribution compared to that of thermal inertia and surface roughness, which is consistent with the findings of \cite{Lagerros97} who investigated the effect using objects random Gaussian shapes. All of these studies justify the use of convex, prolate ellipsoid studies in our TPM approach.

Although not the case here, thermal inertia bias could arise if the size is fixed during the fitting procedure. For example, diameters that are estimated from radar observations can suffer from an overestimated z-axis if the object is observed at near-equatorial view, yielding a higher thermal inertia \citep{Rozitis&Green14}. Some studies have investigated how changes in shape and spin vector can affect the value of thermal inertia. Using both a spherical and radar shape model for the asteroid (101955) Bennu, \cite{Emery_etal14} found that the radar shape model gave a lower thermal inertia result compared to a spherical shape with the same spin vector. The lower thermal inertia is explained by the oblateness of Bennu, as surface facets are systematically tilted away from the sun-direction, cooling the surface temperatures relative to that of a sphere and requiring a lower model thermal inertia to compensate. As briefly mentioned before, our use of prolate ellipsoids may likely be the cause of overestimation of the elongation of objects. For example: if we were to make b/c = 1.1, the average orientation of the shape facets would veer away from the sun and effectively lower the modeled surface temperatures. By lengthening the b/c axis, there is less need to lengthen the a/b axis in order to match the surface temperature.

It is becoming increasingly common for thermal infrared observations to be used to refine the detailed shape models created from the delay-Doppler radar and/or the visible lightcurve inversion techniques \cite{Durech_etal17}. Shape models from these methodologies can sometimes yield incorrect z-axis if the object is observed nearly equator-on \citep{Rozitis_etal13,Rozitis&Green14}. In short, thermal data constrains the effective diameter and by extension the z-axis dimension if the x- \& y-axis dimensions were known to great accuracy from the observations. Additionally, \cite{Hanus_etal15} utilized thermal infrared data to refine convex shape models derived from lightcurve inversion. They varied a convex shape model within the uncertainty of the photometric errors to generate a set of shapes to then be used to generate thermal fluxes from a TPM which were then fit to WISE observations. Although not much work has been done to derive new shapes from only disk-integrated thermal emission data, approaches that combine thermal observations with optical lightcurves \citep{Durech_etal12,Durech_etal14} and other telescopic observations \citep[e.g. stellar occultations, optical interferometry and delay-Doppler radar;][]{Durech_etal15,Durech_etal17} have been employed in order to refine pre-existing shape models. Generally, the refined shape models appear smoother \citep{Durech_etal12,Hanus_etal15}, which is likely in part due to the thermal emission being sensitive to large-scale curvature, especially at lower thermal inertia values \cite{Lagerros96}.


\section{Application to WISE Observations}\label{sec5}

In this section we apply our multi-epoch TPM approach to WISE observations of asteroids that were used in our synthetic dataset. The TPM implimentation on the WISE data is the same as described in \cref{subsec:33} except that we incorporate surface roughness in order to account for surface topography effects. We step through three default roughness values ($\bar{\theta} = {10^\circ,\ 29^\circ,\ 58^\circ}$) that \cite{Delbo&Tanga09} used for IRAS observations.

\subsection{Data Description}

The WISE mission, an astrophysics mission designed to map the entire sky, operated in its fully cryogenic mode from 14 January to 5 August in 2010 at wavelengths centered near 3.4, 4.6, 12 and 22 $\mu m$, denoted W1, W2, W3 \& W4 \citep{Wright_etal10}. A data-processing enhancement called NEOWISE \citep{Mainzer_etal11a} detected moving Solar System objects, most of which were asteroids in the main-belt and in near-Earth orbits. During each grouping of observations (epoch), a moving object is typically detected around 10 to 20 times, in $\sim$1.6 hr multiples - the orbital period of the spacecraft. This means that flux measurements are separated in time by more than $\sim$1.6 hr. Therefore, depending on the object's range of motion on the sky, each epoch of observations can potentially span up to 36 hr. NEOWISE reports each moving object detection to the Minor Planet Center (MPC\footnote{\url{http://www.minorplanetcenter.net/}}), where the start time, RA and Dec of each observation can be retrieved. The set of  times and locations are used to parse the WISE All-Sky Single Exposure (L1b) catalog on the Infrared Science Archive (IRSA) maintained by the Infrared Science and Analysis Center (IPAC\footnote{\url{http://irsa.ipac.caltech.edu/Missions/wise.html}}). We select detections reported to within 10$''$ and $10$ s of those reported to the MPC. These constraints are ``relaxed'' relative to the accuracy of the telescope's astrometric precision to guarantee that IPAC returns flux information for each reported MPC observation. These criteria also return many spurious detections. We discuss in the following our method of rejecting spurious sources returned from these generous search criteria.

\doublespacing
\begin{footnotesize}
\begingroup
\setlength{\LTleft}{-20cm plus -1fill}
\setlength{\LTright}{\LTleft}
\begin{longtable}[t!]{lrcrccrcccc}
\caption{WISE Observation Circumstances and Fluxes}\label{table2}\\

Object & UT Date \footnotemark[1] & $\Delta t_\textrm{obs}$\footnotemark[2] & $N_\textrm{obs}$\footnotemark[3] & $R_\textrm{AU}$\footnotemark[4] & $\Delta_\textrm{AU}$\footnotemark[5] & \multicolumn{1}{c}{$\alpha$ ($^\circ$)\footnotemark[6]} & $\overline{W3}$\footnotemark[7] & $\diamondsuit{W3}$\footnotemark[8] & $\overline{W4}$\footnotemark[7] & $\diamondsuit{W4}$\footnotemark[8] \\ \hline \multicolumn{7}{c}{ } \\[-1,6em] \hline \endfirsthead

\multicolumn{11}{c}{\tablename\ \thetable{} -- continued}\\
\vspace{-0.25cm}\\ 
Object & UT Date\footnotemark[1] & $\Delta t_\textrm{obs}$\footnotemark[2] & $N_\textrm{obs}$\footnotemark[3] & $R_\textrm{AU}$\footnotemark[4] & $\Delta_\textrm{AU}$\footnotemark[5] & \multicolumn{1}{c}{$\alpha$ ($^\circ$)\footnotemark[6]} & $\overline{W3}$\footnotemark[7] & $\diamondsuit{W3}$\footnotemark[8] & $\overline{W4}$\footnotemark[7] & $\diamondsuit{W4}$\footnotemark[8] \\  \hline \multicolumn{7}{c}{ } \\[-1,6em] \hline \endhead

\hline
\multicolumn{11}{l}{All mean flux and range values are in units of mJy = $10^{-29}$ Wm$^{-2}$ Hz$^{-1}$.}\\
\multicolumn{11}{l}{$^1${UT date of the first observation}}\\
\multicolumn{11}{l}{$^2${Time spanned by observations (days)}}\\
\multicolumn{11}{l}{$^3${Number of observations used}}\\
\multicolumn{11}{l}{$^4${Mean Heliocentric distance}}\\
\multicolumn{11}{l}{$^5${Mean WISE-centric distance}}\\
\multicolumn{11}{l}{$^6${Mean solar phase angle}}\\
\multicolumn{11}{l}{$^7${Lightcurve-averaged mean flux}}\\
\multicolumn{11}{l}{$^8${Photometric range of lightcurve}}
\endlastfoot

\vspace{-0.25cm} & 9 Feb 2010 & 1.257 & 13 & 2.840 & 2.647 & 20.33 & 451.8 $\pm$ 5.2 &  141.6 $\pm$ 10.3 & 1224 $\pm$ 26 & 350.2 $\pm$ 65.8 \\
\vspace{-0.25cm}(167) Urda \\
 & 2 Aug 2010 & 1.257 & 12 & 2.787 & 2.510 & -21.27 & 601.1  $\pm$ 7.2 & 185.9 $\pm$ 13.8 & 1559 $\pm$ 30 & 480.7 $\pm$  51.1 \\
\hline
\vspace{-0.25cm} & 4 Feb 2010 & 3.772 & 22 & 3.718 & 3.580 & 15.38 & 87.65 $\pm$ 1.27 & 36.88 $\pm$ 2.57 & 322.7 $\pm$ 9.0 & 119.9 $\pm$ 16.1 \\
\vspace{-0.25cm}(183) Istria \\
 & 21 Jul 2010 & 1.257 & 13 & 3.765 & 3.562 & -15.62 & 70.84 $\pm$ 1.19 & 43.98 $\pm$ 2.23 & 274.7 $\pm$ 8.0 & 160.0 $\pm$ 14.9 \\
\hline
\vspace{-0.25cm} & 10 Feb 2010 & 0.595 & 9 & 2.888 & 2.700 & 19.98 & 508.2 $\pm$ 5.7 & 119.8 $\pm$ 10.6 & 1358 $\pm$ 28 & 344.8 $\pm$ 57.3 \\
\vspace{-0.25cm}(208) Lacrimosa \\
 & 8 Aug 2010 & 1.257 & 10 & 2.913 & 2.645 & -20.30 & 442.4 $\pm$ 4.9 & 97.78 $\pm$ 9.89 & 1269 $\pm$ 25 & 255.1 $\pm$ 48.3 \\
\hline
\vspace{-0.25cm} & 10 Feb 2010 & 1.257 & 14 & 3.180 & 3.009 & 18.08 & 164.9 $\pm$ 2.2 & 72.85 $\pm$ 3.98 & 534.7 $\pm$ 11.5 & 208.9 $\pm$ 24.1 \\
\vspace{-0.25cm}(413) Edburga \\
 & 26 Jul 2010 & 1.257 & 15 & 2.701 & 2.427 & -22.01 & 478.4 $\pm$ 5.8 & 180.3 $\pm$ 11.6 & 1219 $\pm$ 25 & 410.4 $\pm$ 42.0 \\
\hline
\vspace{-0.25cm} & 18 Jan 2010 & 0.596 & 8 & 3.342 & 3.164 & 17.11 & 440.3 $\pm$ 5.1 & 230.6 $\pm$ 9.9 & 1405 $\pm$ 30 & 594.3 $\pm$ 67.6 \\
\vspace{-0.25cm}(509) Iolanda \\
 & 3 Jul 2010 & 1.257 & 15 & 3.327 & 3.080 & -17.72 & 470.5 $\pm$ 6.0 & 216.4 $\pm$ 12.2 & 1461 $\pm$ 24 & 649.9 $\pm$ 40.7 \\
\hline
\vspace{-0.25cm} & 29 Jan 2010 & 1.257 & 10 & 2.764 & 2.593 & 20.88 & 223.7 $\pm$ 2.8 & 74.50 $\pm$ 5.24 & 618.4 $\pm$ 15.4 & 223.4 $\pm$ 28.7 \\
\vspace{-0.25cm}(771) Libera \\
 & 16 Jul 2010 & 3.903 & 22 & 3.111 & 2.847 & -18.97 & 146.2 $\pm$ 2.1 & 75.75 $\pm$ 4.31 & 474.6 $\pm$ 11.2 & 223.2 $\pm$ 19.7 \\
\hline
\vspace{-0.25cm} & 19 Jan 2010 & 1.125 & 14 & 2.329 & 2.085 & 24.99 & 180.8 $\pm$ 2.4 & 70.40 $\pm$ 5.26 & 373.4 $\pm$ 9.3 & 127.4 $\pm$ 23.6 \\
\vspace{-0.25cm}(857) Glasenappia \\
 & 11 Jul 2010 & 1.389 & 17 & 2.172 & 1.823 & -27.75 & 279.2 $\pm$ 3.5 & 80.68 $\pm$ 6.9 & 527.0 $\pm$ 12.7 & 164.6 $\pm$ 21.7 \\
\hline
\vspace{-0.25cm} & 31 Jan 2010 & 1.125 & 9 & 3.324 & 3.188 & 17.24 & 161.3 $\pm$ 2.2 & 76.53 $\pm$ 4.46 & 531.1 $\pm$ 10.99 & 214.3 $\pm$ 19.85 \\
\vspace{-0.25cm}(984) Gretia \\
 & 21 Jul 2010 & 1.125 & 10 & 3.159 & 2.920 & -18.71 & 192.6 $\pm$ 2.5 & 81.18 $\pm$ 5.08 & 604.3 $\pm$ 14.8 & 258.7 $\pm$ 32.8 \\
\hline
\vspace{-0.25cm} & 15 Jan 2010 & 0.860 & 9 & 3.897 & 3.729 & 14.61 & 72.50 $\pm$ 1.10 & 12.50 $\pm$ 2.16 & 294.7 $\pm$ 7.0 & 39.25 $\pm$ 14.02 \\
\vspace{-0.25cm}(1036) Ganymed \\
 & 22 Jun 2010 & 1.125 & 14 & 3.463 & 3.241 & -17.02 & 141.3 $\pm$ 2.1 & 27.94 $\pm$ 4.14 & 494.8 $\pm$ 12.1 & 81.66 $\pm$ 24.09 \\
\hline
\vspace{-0.25cm} & 13 Feb 2010 & 1.125 & 11 & 3.075 & 2.900 & 18.73 & 184.4 $\pm$ 2.4 & 77.33 $\pm$ 4.85 & 524.7 $\pm$ 12.0 & 201.6 $\pm$ 26.8 \\
\vspace{-0.25cm}(1140) Crimea \\
 & 1 Aug 2010 & 1.124 & 11 & 2.990 & 2.725 & -19.76 & 220.7 $\pm$ 2.9 & 110.0 $\pm$ 5.1 & 624.5 $\pm$ 12.6 & 290.7 $\pm$ 24.4 \\
\hline
\vspace{-0.25cm} & 18 Jan 2010 & 0.992 & 10 & 2.580 & 2.354 & 22.41 & 59.19 $\pm$ 1.11 & 41.00 $\pm$ 2.05 & 158.7 $\pm$ 6.3 & 99.88 $\pm$ 12.78 \\
\vspace{-0.25cm}(1188) Gothlandia \\
 & 2 Jul 2010 & 1.389 & 16 & 2.521 & 2.222 & -23.68 & 88.37 $\pm$ 1.40 & 58.34 $\pm$ 2.47 & 214.4 $\pm$ 6.8 & 129.9 $\pm$ 11.8 \\
\hline
\vspace{-0.25cm} & 21 Jan 2010 & 0.992 & 8 & 3.210 & 3.038 & 17.85 & 117.5 $\pm$ 1.8 & 122.0 $\pm$ 3.6 & 367.8 $\pm$ 9.7 & 350.8 $\pm$ 18.2 \\
\vspace{-0.25cm}(1291) Phryne \\
 & 8 Jul 2010 & 1.257 & 14 & 3.090 & 2.825 & -19.12 & 131.0 $\pm$ 2.1 & 142.4 $\pm$ 4.3 & 393.8 $\pm$ 11.4 & 386.3 $\pm$ 18.9 \\
\hline
\vspace{-0.25cm} & 4 Feb 2010 & 3.771 & 23 & 2.699 & 2.511 & 21.43 & 15.44 $\pm$ 0.60 & 4.102 $\pm$ 1.217 & 43.91 $\pm$ 3.48 & 17.13 $\pm$ 8.23 \\
\vspace{-0.25cm}(1432) Ethiopia \\
 & 28 Jul 2010 & 1.389 & 15 & 2.307 & 1.992 & -26.02 & 32.16 $\pm$ 0.76 & 7.282 $\pm$ 1.496 & 78.34 $\pm$ 3.84 & 14.90 $\pm$ 7.49 \\
\hline
\vspace{-0.25cm} & 8 Feb 2010 & 4.301 & 19 & 2.490 & 2.284 & 23.33 & 90.81 $\pm$ 1.46 & 42.36 $\pm$ 3.0 & 208.3 $\pm$ 6.6 & 81.53 $\pm$ 12.85 \\
\vspace{-0.25cm}(1495) Helsinki \\
 & 5 Aug 2010 & 1.125 & 10 & 2.267 & 1.940 & -26.47 & 140.6 $\pm$ 2.0 & 87.26 $\pm$ 4.38 & 299.8 $\pm$ 8.3 & 154.2 $\pm$ 16.5 \\
\hline
\vspace{-0.25cm} & 24 Jan 2010 & 1.125 & 12 & 2.950 & 2.776 & 19.50 & 27.39 $\pm$ 0.68 & 14.91 $\pm$ 1.37 & 82.59 $\pm$ 3.85 & 51.08 $\pm$ 7.46 \\
\vspace{-0.25cm}(1568) Aisleen \\
 & 7 Jul 2010 & 1.257 & 14 & 3.100 & 2.550 & -21.04 & 52.16 $\pm$ 0.90 & 21.74 $\pm$ 1.84 & 135.1 $\pm$ 4.8 & 63.58 $\pm$ 9.57 \\
\hline
\vspace{-0.25cm} & 21 Jan 2010 & 0.993 & 10 & 3.285 & 3.116 & 17.43 & 28.51 $\pm$ 0.71 & 20.45 $\pm$ 1.44 & 94.41 $\pm$ 4.53 & 63.07 $\pm$ 8.75 \\
\vspace{-0.25cm}(1607) Mavis \\
 & 4 Jul 2010 & 1.125 & 12 & 3.039 & 2.778 & -19.47 & 35.16 $\pm$ 0.73 & 14.46 $\pm$ 1.44 & 107.9 $\pm$ 3.9 & 48.81 $\pm$ 8.31 \\
\hline
\vspace{-0.25cm} & 23 Jan 2010 & 0.992 & 11 & 2.333 & 2.109 & 24.96 & 13.79 $\pm$ 0.53 & 5.748 $\pm$ 1.041 & 34.52 $\pm$ 2.83 & 22.68 $\pm$ 5.61 \\
\vspace{-0.25cm}(1980) Tezcatlipoca \\
 & 30 Jun 2010 & 1.257 & 8 & 2.064 & 1.716 & -29.39 & 29.65 $\pm$ 0.73 & 18.61 $\pm$ 1.54 & 65.74 $\pm$ 3.46 & 42.50 $\pm$ 6.61 \\
\hline
\vspace{-0.25cm} & 25 Jan 2010 & 0.992 & 9 & 2.688 & 2.502 & 21.49 & 23.82 $\pm$ 0.68 & 16.42 $\pm$ 1.38 & 64.43 $\pm$ 3.85 & 34.34 $\pm$ 8.27 \\
\vspace{-0.25cm}(2156) Kate \\
 & 11 Jul 2010 & 1.126 & 11 & 2.625 & 2.322 & -22.67 & 24.93 $\pm$ 0.66 & 15.41 $\pm$ 1.29 & 66.16 $\pm$ 3.66 & 37.33 $\pm$ 7.50 \\
\hline
\vspace{-0.25cm} & 25 Jan 2010 & 0.993 & 10 & 3.103 & 2.944 & 18.50 & 21.82 $\pm$ 0.66 & 10.80 $\pm$ 1.36 & 72.11 $\pm$ 3.44 & 31.03 $\pm$ 7.05 \\
\vspace{-0.25cm}(4611) Vulkaneifel \\
 & 11 Jul 2010 & 1.257 & 13 & 3.100 & 2.834 & -19.05 & 25.08 $\pm$ 0.65 & 13.63 $\pm$ 1.25 & 81.09 $\pm$ 3.86 & 40.84 $\pm$ 7.26 \\
\hline
\vspace{-0.25cm} & 28 Jan 2010 & 0.993 & 11 & 3.180 & 3.032 & 18.04 & 28.80 $\pm$ 0.72 & 19.98 $\pm$ 1.43 & 91.72 $\pm$ 4.18 & 49.67 $\pm$ 8.86 \\
\vspace{-0.25cm}(5625) 1991 AO$_2$ \\
 & 14 Jul 2010 & 1.125 & 13 & 3.168 & 2.900 & -18.61 & 34.55 $\pm$ 0.83 & 31.91 $\pm$ 1.63 & 107.8 $\pm$ 4.3 & 100.6 $\pm$ 7.7 \\
\hline
\vspace{-0.25cm} & 3 Feb 2010 & 4.301 & 19 & 2.341 & 2.124 & 24.91 & 13.21 $\pm$ 0.84 & 11.16 $\pm$ 1.49 & 31.40 $\pm$ 3.64 & 27.42 $\pm$ 7.01 \\
\vspace{-0.25cm}(6159) 1991 YH \\
 & 30 Jul 2010 & 1.389 & 11 & 2.424 & 2.120 & -24.67 & 14.90 $\pm$ 0.59 & 11.15 $\pm$ 1.17 & 36.87 $\pm$ 3.36 & 27.63 $\pm$ 6.68 \\

\end{longtable}
\endgroup
\end{footnotesize}
\singlespacing

According to the WISE Explanatory Supplement \citep{Cutri_etal12} photometric profile fits are unreliable for W3 $<$ -3 mag or W4 $<$ -4 mag, due to saturation of the detector. Since non-linearity is present for sources with W3 $<$ 3.6 mag and W4 $<$ -0.6 mag, we increase the magnitude uncertainty to 0.2 mag for objects brighter than these values \citep{Mainzer_etal11b}.
We shift the isophotal wavelengths and zero magnitude point of W3 \& W4 to account for the red-blue calibrator discrepancy described in the Explanatory Supplement. The raw magnitudes that are reported were calibrated assuming that the flux across each filter was that of Vega's spectrum. Thus a color-correction must be made to account for the discrepancy between the spectrum of the object to that of Vega \citep{Wright_etal10,Cutri_etal12}. For each individual observation, NEATM was used to calculate the flux spectrum across the full bandpass for each WISE band for use in the color correction, which ranged from 0.87 to 1.0 for W3 and was nearly constant at 0.98 for W4. Based on an analysis of asteroid flux uncertainties in consecutive frames by \cite{Hanus_etal15}, we increase the flux error in W3 \& W4 by a factor of 1.4 and 1.3, respectively.

In order to filter out bad observations of an asteroid, for example in a situation in which it passes near a background star or when the query returns a detection of an unwanted object, we employ ``Peirce's criterion''\footnote{This procedure for rejecting outlier data points uses a criterion based on Gaussian statistics. In short, rejection of a data point occurs when the probability of the deviation from the mean obtained by retaining the data in question is less than that of the deviation from the mean obtained by their rejection, multiplied by the probability of making as many, and no more, outlier observations. The motivation for using this relatively obscure procedure is due to the fact that it makes no arbitary assumptions about the cutoff for outliers {\it and} can be used to simultaneously identify multiple outliers. Peirce's criterion is rigorous and generalized in its applicability, when compared to William Chauvenet's criterion \citep{Taylor97}.} \citep{Peirce1852,Gould1855} as outlined and demonstrated by \cite{Ross03}. We flag spurious observations using this algorithm based on W4--W3, since it will identify and remove sources of an anomalous color temperature. Using color, rather than raw flux, avoids the possibility of removing seemingly anomalous observations of the minimum or maximum flux of a highly elongated object. This rejection method is a simpler alternative to that implemented by \cite{AliLagoa_etal13,Hanus_etal15,Hanus_etal18} on WISE data, in which the WISE inertial source catalog was checked explicitly for possible flux contamination of stars.

Mean fluxes, denoted as $\overline{W3}$ \& $\overline{W4}$, were calculated by simply taking the error-weighted mean of all observations. The photometric range for each band, denoted as $\diamondsuit{W3}$ \& $\diamondsuit{W4}$, was calculated by subtracting the minimum and maximum fluxes. Standard error propagation was used to estimate 1-$\sigma$ uncertainties for each of these parameters, and input as $\sigma_o$ in \cref{eq11}. These bright, slow-moving objects with $P_\textrm{rot} < $ 36 hr have sufficient coverage in rotational phase to provide estimates of these parameters. However, some objects may have only been detected handful of times, which could only sparsely sample the rotational phases. In a follow-up paper we will develop and present a more rigorous method for accurately estimating the flux mean and range, with associated errors, from sparse lightcurve coverage.

\doublespacing
\begin{footnotesize}
\begingroup
\setlength{\LTleft}{-20cm plus -1fill}
\setlength{\LTright}{\LTleft}
\begin{longtable}{lcccccccc}
\caption{WISE Data TPM Results}\label{table3}\\

Object & $D_\mathit{eff}$ (km) & $p_{V}$ & $\Gamma$* & $\bar{\theta} (^{\circ}) $ & a/b$^1$ & Spin$^2$ & $\tilde{\chi}_\mathit{min}^2$ & MBA/NEA \\ \hline \multicolumn{9}{c}{ } \\[-1,6em] \hline \endfirsthead

\multicolumn{9}{c}{\tablename\ \thetable{} -- continued}\\
\vspace{-0.25cm}\\ 
Object & $D_\mathit{eff}$ (km) & $p_{V}$ & $\Gamma$* & $\bar{\theta} (^{\circ}) $ & a/b$^1$ & Spin$^2$ & $\tilde{\chi}_{min}^2$ & MBA/NEA \\  \hline \multicolumn{9}{c}{ } \\[-1,6em] \hline \endhead

\hline
\endfoot

\hline
\multicolumn{9}{l}{*Thermal inertia values are in SI units (Jm$^{-2}$K$^{-1}$s$^{-1/2}$).}\\
\multicolumn{9}{l}{$^1$a/b values are adjusted downward by 16\% to account for the over-estimation as described in \cref{subsec:42}.}\\
\multicolumn{9}{l}{$^2$Indicates either prograde ($\Uparrow$) or retrograde ($\Downarrow$) spin direction.}\\
\multicolumn{9}{l}{$^\dagger$TPM results with $\chi^2_\mathit{min} > 8$ and thus should be used with caution.}
\endlastfoot

(167) Urda & 39.48 $\pm$ 0.89 & $0.252^{+0.010}_{-0.019}$ & $51^{+20}_{-16}$ & 38 $\pm$ 13 & 1.51 $\pm$ 0.27  & $\Downarrow$ & 1.27 & MBA \\
(183) Istria & 31.43 $\pm$ 2.92 & $0.288^{+0.029}_{-0.033}$ & $21^{+12}_{-10}$ & 47 $\pm$ 13 & 1.51 $\pm$ 0.27 & $\Uparrow$ & 5.16 & MBA \\
(208) Lacrimosa & 40.44 $\pm$ 1.37 & $0.253^{+0.012}_{-0.014}$ & $77^{+31}_{-22}$ & 41 $\pm$ 13 & 1.51 $\pm$ 0.27 & $\Uparrow$ & 1.66 & MBA \\
(413) Edburga & 33.44 $\pm$ 1.75 & $0.169^{+0.012}_{-0.022}$ & $41^{+19}_{-10}$ & 9 $\pm$ 6 & 1.51 $\pm$ 0.27 & $\Downarrow$ & 1.98 & MBA \\
(509) Iolanda & 54.39 $\pm$ 3.86 & $0.243^{+0.019}_{-0.027}$ & $8.6^{+12.2}_{-8.6}$ & 18 $\pm$ 7 & 1.51 $\pm$ 0.27 & $\Uparrow$ & 3.01 & MBA \\
(771) Libera$^\dagger$ & 29.23 $\pm$ 2.10 & $0.160^{+0.013}_{-0.019}$ & $61^{+34}_{-26}$ & 53 $\pm$ 39 & 1.51 $\pm$ 0.27 & $\Downarrow$ & 13.5 & MBA \\
(857) Glasenappia & 13.62 $\pm$ 0.84 & $0.297^{+0.013}_{-0.021}$ & $58^{+30}_{-24}$ & $< 20$ & 2.16 $\pm$ 0.39 & $\Uparrow$ & 1.52 & MBA \\
(984) Gretia & 34.72 $\pm$ 1.18 & $0.227^{+0.012}_{-0.023}$ & $28^{+8}_{-7}$ & 53 $\pm$ 10 & 1.51 $\pm$ 0.27 & $\Uparrow$ & 1.81 & MBA \\
(1036) Ganymed & 35.85 $\pm$ 1.95 & $0.278^{+0.017}_{-0.027}$ & $15 ^{+22}_{-15}$ & 40 $\pm$ 32 & 1.08 $\pm$ 0.19 & | & 0.55 & NEA \\
(1140) Crimea & 30.13 $\pm$ 1.18 & $0.276^{+0.020}_{-0.028}$ & $23^{+15}_{-23}$ & 23 $\pm$ 21 & 1.51 $\pm$ 0.27 & $\Downarrow$ & 1.50 & MBA \\
(1188) Gothlandia & 13.52 $\pm$ 0.84 & $0.238^{+0.019}_{-0.022}$ & $38^{+21}_{-13}$ & 41 $\pm$ 27 & 1.51 $\pm$ 0.27 & $\Downarrow$ & 0.43 & MBA \\
(1291) Phryne & 27.03 $\pm$ 1.65 & $0.186^{+0.014}_{-0.017}$ & $20^{+16}_{-6}$ & 45 $\pm$ 17 & 2.16 $\pm$ 0.39 & $\Uparrow$ & 1.78 & MBA \\
(1432) Ethiopia & \ 7.15 $\pm$ 0.67 & $0.535^{+0.058}_{-0.070}$ & $71^{+180}_{-65}$ & | & 1.51 $\pm$ 0.27 & $\Uparrow$ & 0.56 & MBA \\
(1495) Helsinki & 13.31 $\pm$ 0.59 & $0.271^{+0.017}_{-0.033}$ & $19^{+13}_{-13}$ & 12 $\pm$ 8 & 1.51 $\pm$ 0.27 & $\Uparrow$ & 0.52 & MBA \\
(1568) Aisleen & 11.66 $\pm$ 1.01 & $0.328^{+0.034}_{-0.038}$ & $51^{+41}_{-22}$ & 46 $\pm$ 38 & 2.16 $\pm$ 0.39 & $\Downarrow$ & 2.74 & MBA \\
(1607) Mavis$^\dagger$ & 14.52 $\pm$ 1.72 & $0.249^{+0.032}_{-0.037}$ & $37^{+42}_{-25}$ & 58 $\pm$ 50 & 1.51 $\pm$ 0.27 & $\Uparrow$ & 21.45 & MBA \\
(1980) Tezcatlipoca & \ 5.68 $\pm$ 0.58 & $0.205^{+0.035}_{-0.040}$ & $170^{+170}_{-110}$ & 57 $\pm$ 39 & 1.51 $\pm$ 0.27 & $\Downarrow$ & 1.67 & NEA \\
(2156) Kate & \ 8.04 $\pm$ 0.45 & $0.294^{+0.021}_{-0.025}$ & $56^{+23}_{-23}$ & 49 $\pm$ 32 & 2.16 $\pm$ 0.39 & $\Uparrow$ & 0.39 & MBA \\
(4611) Vulkaneifel & 12.10 $\pm$ 1.12 & $0.216^{+0.023}_{-0.028}$ & $32^{+23}_{-32}$ & $< 53$ & 1.51 $\pm$ 0.27 & $\Downarrow$ & 1.01 & MBA \\
(5625) Jamesferguson & 14.46 $\pm$ 0.86 & $0.062^{+0.005}_{-0.006}$ & $52^{+14}_{-15}$ & $> 36$ & 2.16 $\pm$ 0.39 & $\Downarrow$ & 2.46 & MBA \\
(6159) Andreseloy & \ 5.65 $\pm$ 1.37 & $0.247^{+0.061}_{-0.061}$ & $60^{+177}_{-60}$ & | & 1.51 $\pm$ 0.27 & $\Downarrow$ & 0.19 & MBA \\

\end{longtable}
\endgroup
\end{footnotesize}
\singlespacing

\subsection{Results and Discussion}

A summary of our TPM fits and corresponding 1-$\sigma$ uncertainties are given in \cref{table3}. For diameter, albedo and thermal inertia, uncertainties are based on the $\chi^2$ values calculated during the fitting routine. In some cases two values of roughness could not be distinguished as better than the other, so both values are reported. In the case of (6159) Andreseloy all roughness values tried provided statistically indistinguishable fits to the data. The spin direction reported for each object here reflects the preferred spin vector. Often, many spin vector solutions lie within the 1-$\sigma$ uncertainty bounds, thus reporting a single solution would not be meaningful. The spin direction of (1036) Ganymed could not be independently determined here, which is likely due to the nearly-spherical shape of the object combined with its low thermal inertia reducing the asymmetry of the thermal phase curve, as demonstrated in \cref{fig1}(b).

Comparing the diameter estimates from NEATM fits presented by the WISE team \citep[i.e.][]{Mainzer_etal11b,Masiero_etal11} to the values obtained here, we observe agreement (within $\pm$ 15\% of another) between the two sets of results. Yet, for objects $\sim$40 km and above, TPM diameter estimates are systematically higher. The objects in question (Urda, Lacrimosa and Iolanda) did not saturate, nor were they bright enough to lie in the non-linear regime of the WISE detectors. This discrepancy may be due to one the flux corrections described above or from the model differences between our TPM approach and the NEATM used by the WISE team.

When comparing to more previous works, our results are consistent with the thermal inertias reported: (771) Libera and (1980) Tezcatlipoca have thermal inertia estimates of 65 +85/-35 and 220$+380/-204$ Jm$^{-2}$K$^{-1}$s$^{-1/2}$, respectively, made by \citep{Hanus_etal15}, though the reader should take note that the high chi-square indicate that our fits for Libera were relatively poor, contributing to the relatively high parameter uncertainties. (1036) Ganymed has several thermal inertia estimates: 24 $\pm$ 8  \citep{Rozitis_etal18}, 35 +65/-29 \citep{Hanus_etal15}, and 214 $\pm$ 80 Jm$^{-2}$K$^{-1}$s$^{-1/2}$ \citep{Rivkin_etal17}\footnote{Instead of a TPM analysis, this work employed an approach pioneered by \cite{Harris&Drube16} in which the NEATM $\eta$ value is used to indirectly determine the thermal inertia.} The higher estimates are around an order of magnitude greater than the smallest estimates and can be explained by the thermal inertia dependency on surface temperature. Our thermal inertia estimate for Ganymed was based on data collected at $R_\textrm{AU} =$ 3.5 and 3.8 and is thus consistent with similar estimates at the same distance.

\begin{figure}[H]
  \centering
  	\begin{tabular}{c}
    \includegraphics[width=.6\linewidth]{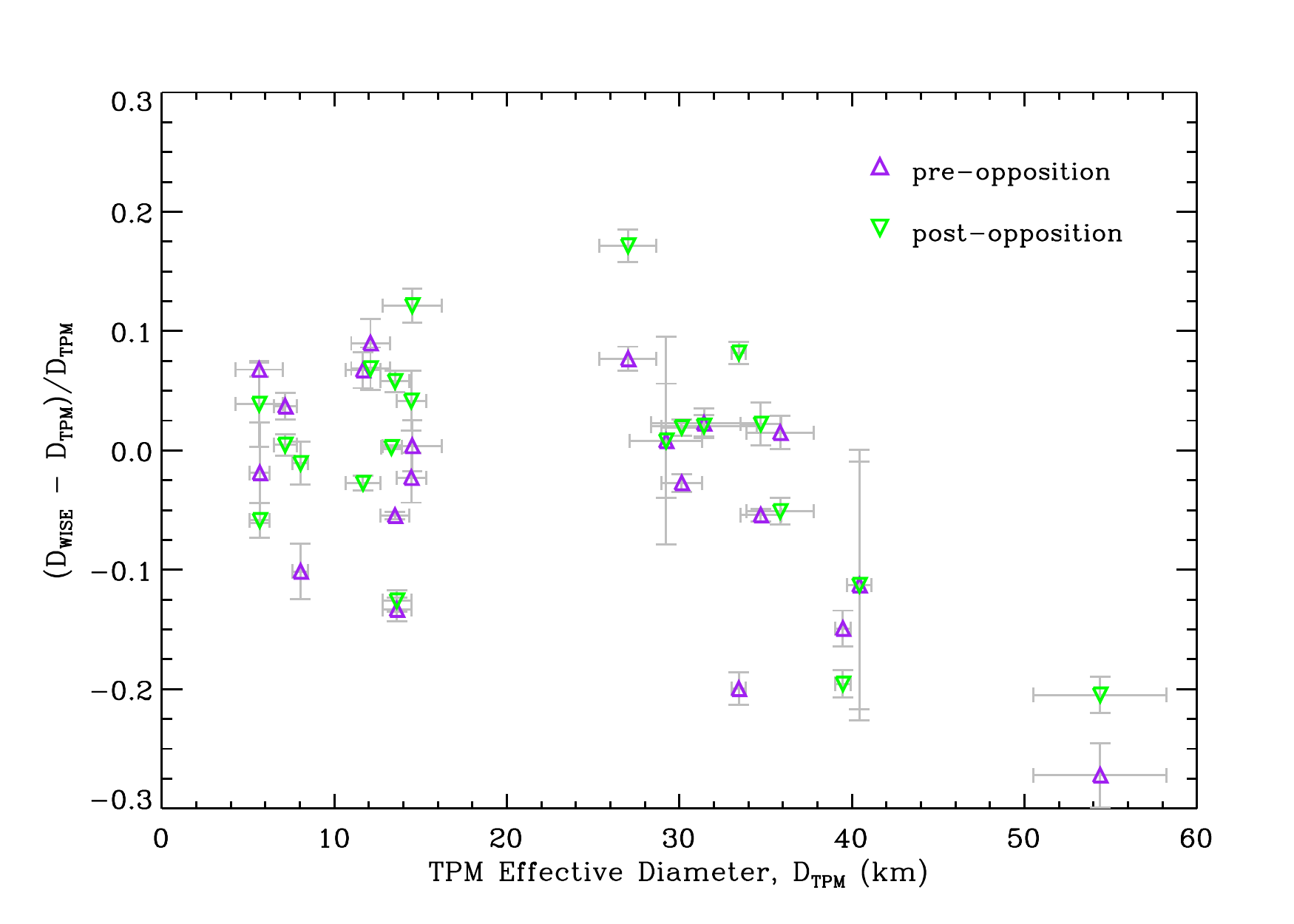} \\
	\includegraphics[width=.6\linewidth]{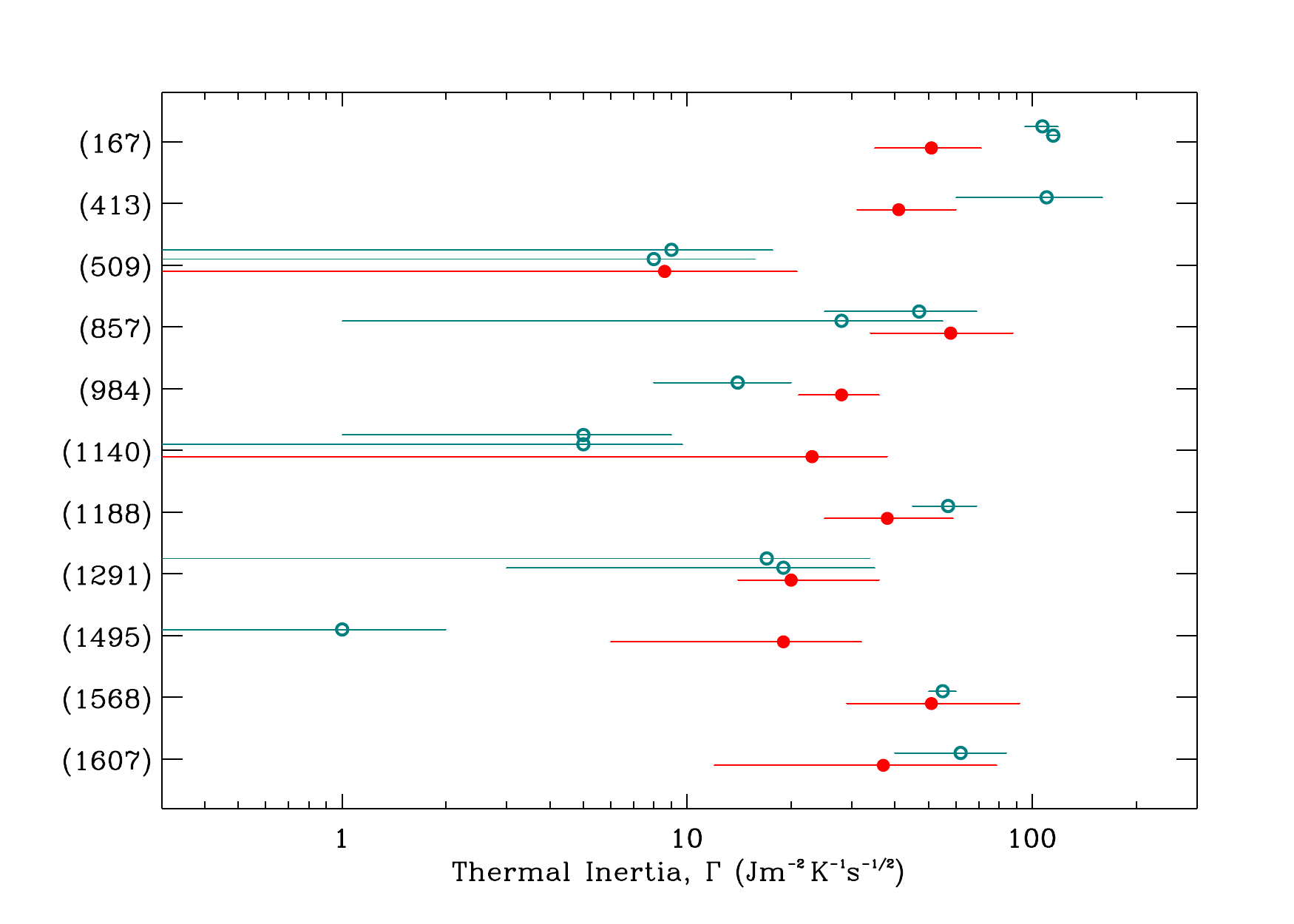}
	\end{tabular}
  \caption{{\it Top}: Comparison of the effective diameter values obtained by \cite{Masiero_etal11} and \cite{Mainzer_etal11b} to our reported TPM values. Purple, upward-facing and green, downward-facing triangles are data collected at pre- and post-opposition, respectively. {\it Bottom}: Comparison of thermal inertia estimates for eleven objects by \cite{Hanus_etal18}, in teal open circles, to ours, in red filled circles.}\label{fig9}
\end{figure}

\begin{figure}[H]
\centering
	\begin{tabular}{c}
	\includegraphics[width=.6\linewidth]{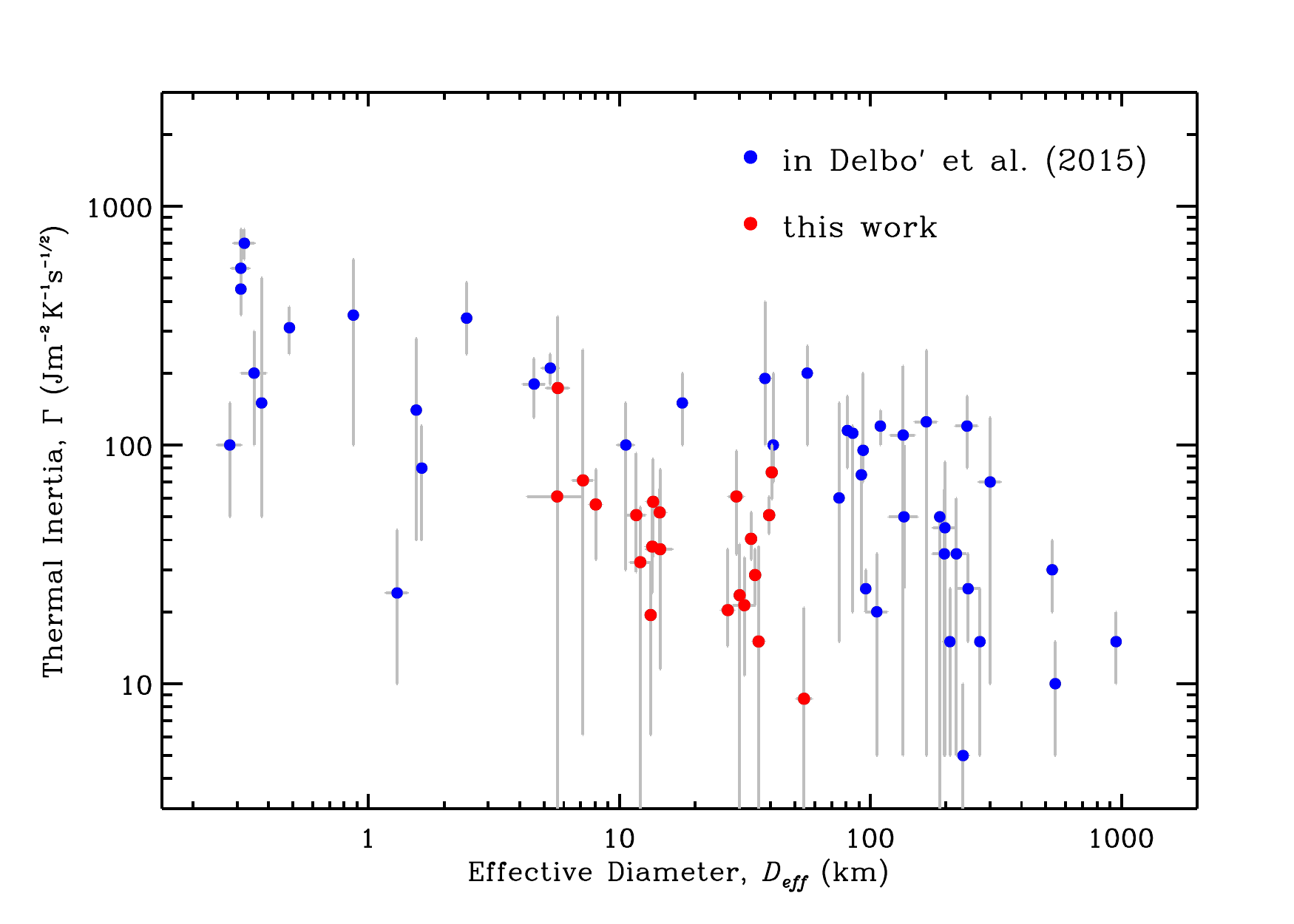} \\
	\includegraphics[width=.6\linewidth]{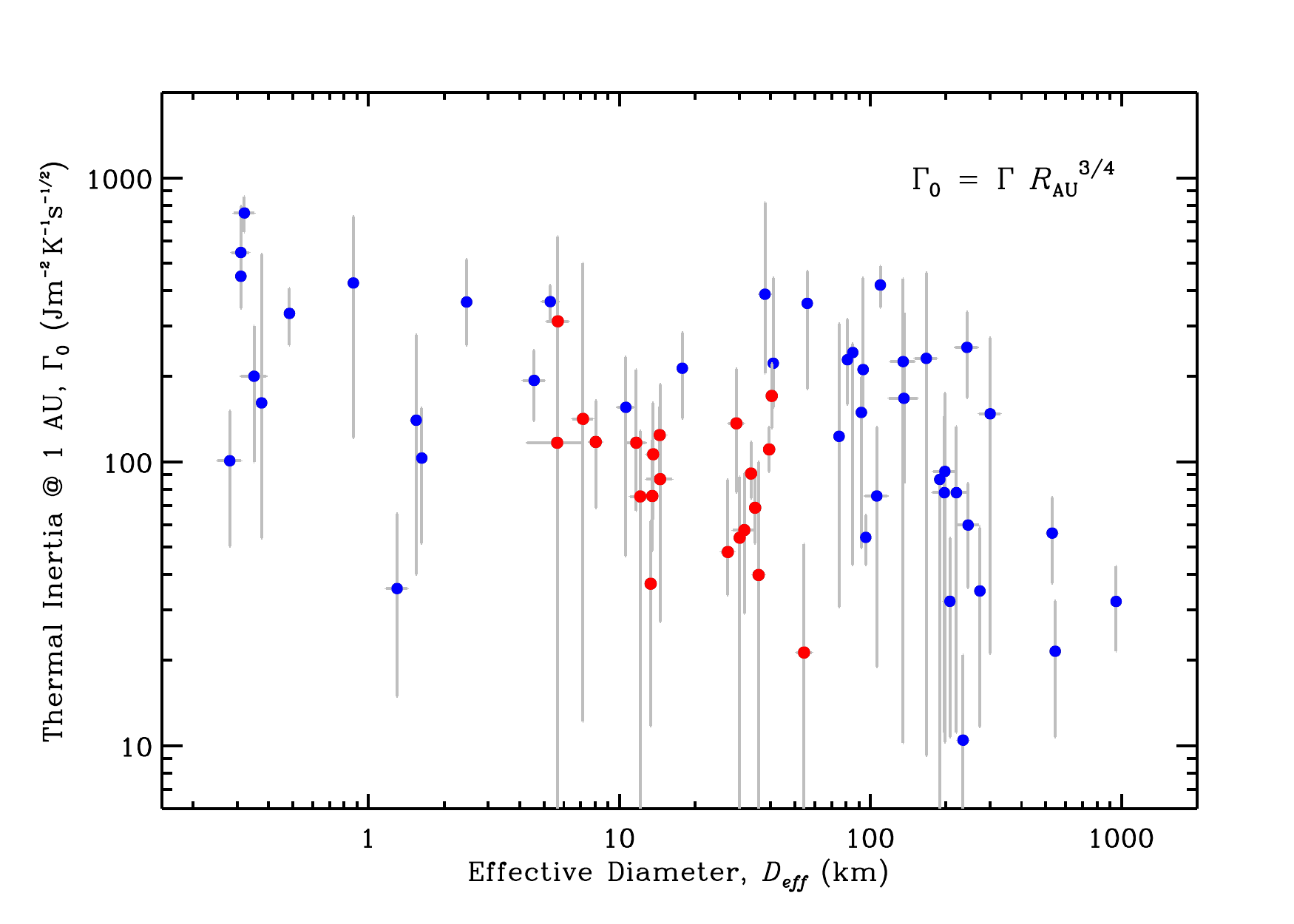}
	\end{tabular}
	\caption{Thermal inertia versus diameter values from this work compared along with previous estimates. The bottom panel shows thermal inertia adjusted for heliocentric distance, as per the text \citep{Rozitis_etal18}.}\label{fig10}
\end{figure}

Eleven objects in this study have been analyzed by the recent work of \cite{Hanus_etal18}, in which the varied-shape TPM was used to derive thermophysical properties from the WISE dataset. A comparison of the thermal inertias for each of these objects in shown in the right panel of \cref{fig9}. Some objects have two estimates presented in \cite{Hanus_etal18} due to ambiguous shape models that produce equivalent fits to the data. For 7 out of these 11 objects, there is very good agreement (i.e. the estimates are within the 1-sigma uncertainties) between the estimates from their work and ours. We note that our model fit for (1607) Mavis was noticeably inaccurate, as indicated by the large $\tilde{\chi}_\mathit{min}^2$ in \cref{table3}, for which we note the error bars for each parameter are noticeably large\footnote{As a reminder, parameter uncertainties are scaled by $\tilde{\chi}_\mathit{min}^2$ values in order to account for best-fit TPM fluxes that deviate from the measured fluxes.} and consistent with \cite{Hanus_etal18}. For two objects, (413) Edburga and (984) Gretia, the 1-sigma error bars just barely miss overlapping and two others, (167) Urda and (1495) Helsinki, have very different estimates. The two of estimates \cite{Hanus_etal18} present for (167) Urda are just over twice as large as ours and have very small reported uncertainties of $\pm 5$. \cite{Hanus_etal18} report a thermal inertia just over zero for (1495) Helsinki, also with a very small uncertainty. For each of these objects, our 2-sigma uncertainty bounds encompass the \cite{Hanus_etal18} estimates and both works overall appear to deliver no systematically different estimates from another.

With the combination of our TPM results with the subset compiled by \cite{Delbo_etal15} and the large dataset of \cite{Hanus_etal18}, we note that the number of known thermal inertias of objects in the 5--50 km size range increases by 20. Restricting our analysis to the compolated thermal inertias from \cite{Delbo_etal15}, we detect a negative correlation with diameter (\cref{fig10}) evidenced by the Spearman's rank coefficient of $r_s$ = -0.55; this correlation remains when including the $\sim$120 thermal inertias of \cite{Hanus_etal18}. From the findings of \cite{Rozitis_etal18}, however, thermal inertia should be adjusted to account for it's dependency on heliocentric distance by using the formula: $\Gamma = \Gamma_\textrm{0} R^a_\textrm{AU}$. Doing this allows for a better comparison of hot, small NEAs and cool, larger MBAs as the effects of temperature are partially accounted for. We use $a = 3/4$ here, which was eariler suggested by \cite{Delbo_etal15} and \cite{Mueller_etal10}. Even when performing the adjustment on asteroid thermal inertias, the correlation remains statistically significant (with $r_s$ = -0.41 and $p < 0.001$) as shown in the bottom panel of \cref{fig10}. However, this correlation becomes insignificant if we use $a = 4/3$, which is large but well within the range of empirically derived values from analyzing individual objects \citep[i.e. Ganymed and 2002 CE$_{26}$;][]{Rozitis_etal18}. While the purpose of this work is not to investigate this dependency, in a follow-up work we plan to increase the number thermal inertia estimates of small main-belt asteroids, combine our results with \cite{Hanus_etal18}, and revisit this dependency in much greater detail.

The prograde/retrograde spin directions reported here are in agreement with those from DAMIT (\cref{table1}) with the exception of three objects: (208) Lacrimosa, (1495) Helsinki, and (6159) Andreseloy. Incorporating the ambiguous spin direction for Ganymed, this means that 17 out of 21 (81 $\pm$ 9\%) objects had matched the spin vectors in the DAMIT shape models. Comparing this result to the findings of our proof-of-concept study in \cref{subsec:42}, there is a noticeable difference. Assuming the DAMIT shape models are 100\% accurate, there is much improvement in the percent of synthetic objects in which the spin direction was correctly identified ($\sim$ 58 and $\sim$ 67\% for ellipsoids and a sphere, respectively). When broken down by thermal inertia, as in panel (f) of \cref{fig3}, these results are comparable to the best-case scenario (spherical shape for $\Gamma = 40$  Jm$^{-2}$K$^{-1}$s$^{-1/2}$), and outperforms each situation in which an ellipsoid shape is used. A reasonable explanation for this discrepancy is the inclusion of roughness in the real-world application of the TPM, which adds to the asymmetry of the thermal phase curve - particularly for asteroids in the main-belt (e.g. \cref{fig1}(b)).

\section{Summary and Future Work}\label{section1.6}

The opportunity to constrain diagnostic thermophysical properties of asteroids by constraining the multi-wavelength phase curves will become more common with the advent of more thermal infrared survey telescopes. We have demonstrated in this paper the accuracy and precision of estimating various asteroid parameters using only pre- and post-opposition multi-wavelength thermal observations for an object, without the aid of a known shape model. By varying the a/b axis of a prolate ellipsoid shape model and sampling from a grid of several spin vectors, unique solutions for diameter/albedo, thermal inertia, and the elongation of an object can be ascertained. A small correction to the best-estimate a/b axis is applied to account for the overestimation of the elongation of the body. Constraints on surface roughness and sense of spin direction are more difficult to obtain, most likely due to the relatively low number of data points compared to the number of free parameters we used in the TPM approach. Additional multi-wavelength observations taken close to opposition would increase the precision of the surface roughness estimate, as the flux beaming effect from surface topography is most pronounced in this configuration. Spin axis estimates would also benefit from additional sets of observations taken at another observing geometry. The strong correlation between the peak-to-trough flux range and the sub-solar latitude can be exploited in this case to increase the accuracy and precision of the spin axis orientation.

The TPM approach outlined int this work was applied to 21 asteroids: 19 MBAs and 2 NEAs (\cref{table3}). Our follow-up paper will feature diameter, albedo, thermal inertia, and roughness estimates for over two-hundred asteroids that were observed by WISE. Results pertaining to surface roughness, shape, and spin sense will also be provided. Future work will focus on investigating and improving upon the accuracy and precision of shape and spin using ellipsoids and multi-epoch thermal observations. With the release of WISE/NEOWISE survey data and large-scale visible (e.g. Pan-STARRS and LSST) surveys it is important that appropriate models are developed alongside the releases of these data sets. In particular, sensitive surveys will discover and observe smaller objects and may be the only available observations for newly discovered, faint asteroids. Thus, models that can make use of these survey observations in order to derive asteroid physical properties will be valuable.

\section*{Acknowledgments}

EMM greatly appreciates the support from the NASA Earth and Space Sciences Fellowship (\# NNX14AP21H). This publication makes use of data products from the Wide-field Infrared Survey Explorer, which is a joint project of the University of California, Los Angeles, and the Jet Propulsion Laboratory/California Institute of Technology, funded by the National Aeronautics and Space Administration.

\appendix
\section{TPM Numerical Techniques}\label{appB}
The parameterized version of the time-dependent, one-dimensional heat diffusion equation is:
\begin{equation}\label{eqB1}
  \frac{\partial T'(z,t')}{\partial t'} = \Theta \frac{\partial^{2} T'(z,t')}{\partial x'^{2}}
\end{equation}
where temperature, time and depth are parameterized into dimensionless form by: $T' = \sfrac{T}{T_{eq}}$, $t' = \sfrac{t \omega}{\Theta}$, $x' = \sfrac{x}{l_{s}}$.
To numerically implement the time-dependent one-dimensional heat diffusion equation we employ the Crank-Nicolson finite-difference approach \citep{Press_etal07}. \Cref{eqB1} is discretized into small finite elements with depth increments of $\delta x'$ and time increments of $\delta t'$. If $T'^{n}_{j}$ is the temperature at time $t'=n\delta t'$ and depth $z=j\delta x'$:
\begin{equation}
  \frac{T'^{n+1}_{j}-T'^{n}_{j}}{\delta t'} = \frac{\Theta}{2} \frac{(T'^{n+1}_{j+1}-2T'^{n+1}_{j}+T'^{n+1}_{j-1})+(T'^{n}_{j+1}-2T'^{n}_{j}+T'^{n}_{j-1})}{(\delta x')^{2}}.
\end{equation}
Grouping the terms by time step gives:
\begin{equation}
  T'^{n}_{j-1} + T'^{n}_{j}\big(\tfrac{2(\delta x')^{2}}{\Theta \delta t'}-2\big) + T'^{n}_{j+1} = -T'^{n+1}_{j-1} + T'^{n+1}_{j}\big(\tfrac{2(\delta x')^{2}}{\Theta \delta t'}+2\big) - T'^{n+1}_{j+1}
\end{equation}
For $D$ depth steps a set of linear equations can be represented by the matrix equation:
\begin{equation}
  \left[ \begin{matrix}
   a &-1 & 0 &\cdots & 0\\
   -1 & a & -1& \ddots & \vdots \\
   0 & \ddots & \ddots & \ddots & 0 \\
   \vdots  & \ddots &-1 &a & -1 \\
   0 &\cdots & 0 & -1 & a  \end{matrix} \right]
\renewcommand{\arraystretch}{1.08}
  \left[ \begin{matrix}
    T'^{n}_0\\
    T'^{n}_1\\
    \vdots\\
    T'^{n}_{D-1}\\
    T'^{n}_D \end{matrix} \right]
 = \left[ \begin{matrix}
    d_0\\
    d_1\\
    \vdots\\
    d_{D-1}\\
    d_D \end{matrix} \right]
\end{equation}
where $a = \tfrac{2(\delta x')^{2}}{\Theta \delta t'} + 2$ and $d_{j} = T'^{n+1}_{j-1} + T'^{n+1}_{j}\big(\tfrac{2(\delta x')^{2}}{\Theta \delta t'}-2 \big) + T'^{n}_{j+1}$ and the upper (surface) and lower boundary conditions implimented as such:
\begin{equation}
  \frac{dT'}{dx'} \bigg|_{surf} = \frac{1}{\delta x'} (T'^{n+1}_{1}-T'^{n+1}_{0})
\end{equation}
\begin{equation}
  T'^{n}_{D}-T'^{n}_{D-1} = 0.
\end{equation}
This system of equations is solved via the \texttt{tridag} routine provided in \cite{Press_etal07}. We choose $\delta x' = l_{s}/5$ and calculate temperatures down to $10\ l_{s}$. The time steps, $\delta t'$, are varied such that for $\Theta =$ 0.055 and 100 there are 1440 and 360 time steps per rotation, respectively. To establish if temperature convergence has been reached for a particular latitude bin we utilize an energy balance criterion to the latest rotation or ensure that the temperatures have not changed substantially from the previous rotation.

\section{Transformation from a Spherical to Ellipsoidal Shape}\label{appC}
A generalized ellipsoid with semi-axes $a \ge b \ge c$ is given in Cartesian coordinates by
\begin{equation}
\frac{x^2}{a^2} + {y^2 \over b^2} + {z^2 \over c^2} -1 = 0,
\end{equation}
and can be parameterized using body-centric coordinates, $\theta$ and $\phi$, in the following way:
\begin{equation}
  x = a \cos(\theta)\cos(\phi)
\end{equation}
\begin{equation}
  y = b \sin(\theta)\cos(\phi)
\end{equation}
\begin{equation}
  z = c \sin(\phi)
\end{equation}
for $\theta \in {[0,2\pi)}$ and $\phi \in {[-{\frac{\pi}{2}}, \frac{\pi}{2}]}$ the body-centric longitude and latitude, respectively. This chosen convention implies that an ellipsoid body rotates around the +c-axis, in the +$\theta$ direction, as per the right-hand rule. The directional derivatives of the radius vector, $\mathbf{r} = \langle x, y, z \rangle$, with respect to the body-centric coordinates can then be used to find the normal vector to the surface at every point:
\begin{equation}
  \textbf{R}_{\theta} \equiv \frac{\partial \textbf{r}}{\partial \theta} = \langle -a \sin(\theta)\cos(\phi), b \cos(\theta)\cos(\phi), 0 \rangle
\end{equation}
\begin{equation}
  \textbf{R}_{\phi} \equiv \frac{\partial \textbf{r}}{\partial \theta} = \langle -a \cos(\theta)\sin(\phi), -b \sin(\theta)\sin(\phi), c\cos(\phi) \rangle
\end{equation}

\begin{equation}
  \textbf{n} = | \textbf{R}_{\theta} \times \textbf{R}_{\phi} | = \cos^{2}(\phi) \langle b c \cos(\theta), a c \sin(\theta), a b \tan(\phi) \rangle
\end{equation}

\begin{equation}
  \| \textbf{n} \| = \cos^{2}{\phi} \, {[\, c^{2}\, (a^{2} \sin^{2}{\phi} + b^{2} \cos^{2}{\theta}) + a^{2} b^{2} \tan^{2}{\phi]}}^{1/2}
\end{equation}
An area element, ($dA$), on the ellipsoid surface can now be expressed as a function of the parameterized coordinates:
\begin{equation}
  dA = \| \textbf{n} \| \, d\theta \, d\phi = \cos^{2}{\phi} \, {[\, c^{2}\, (a^{2} \sin^{2}{\phi} + b^{2} \cos^{2}{\theta} ) + a^{2} b^{2} \tan^{2}{\phi]}}^{1/2} d\theta \, d\phi.
\end{equation}
The surface-normal longitude, $\vartheta$, is calculated by projecting the surface-normal vector onto the $x-y$ plane and then taking the dot product with $\hat{x}$:
\begin{equation}
\cos(\vartheta) = \frac{\cos^{2}{\phi} \, \langle b c \cos{\theta}, a c \sin{\theta}, 0 \rangle}{\cos^{2}{\phi} \, [\, c^{2}\, (a^{2} \sin^{2}{\phi} + b^{2} \cos^{2}{\theta})]^{1/2}}\ \cdot \langle 1, 0, 0 \rangle.
\end{equation}
The surface-normal latitude, $\varphi$, is calculated by taking the dot product of the surface-normal vector and $\hat{z}$̂:
\begin{equation}
\sin(\varphi) = \frac{\cos^{2}{\phi} \langle b c \cos{\theta}, a c \sin{\theta}, a b \tan{\phi} \rangle}{\cos^{2}{\phi} \, {[\, c^{2}\, (a^{2} \sin^{2}{\phi} + b^{2} \cos^{2}{\theta} ) + a^{2} b^{2} \tan^{2}{\phi]}}^{1/2}} \cdot \langle 0, 0, 1 \rangle
\end{equation}
Simplifying, results in two transformation equations for converting body-centric coordinates to surface-normal coordinates:
\begin{equation}
\tan(\vartheta) = \bigg(\frac{a}{b}\bigg) \tan{\theta}
\end{equation}
and
\begin{equation}
\tan(\varphi) = \Bigg(\frac{a b}{c}\Bigg) \frac{\tan(\phi)}{(a^{2} \sin^{2}{\theta} + b^{2} \sin^{2}{\theta})^{1/2}}.
\end{equation}




 \bibliographystyle{elsarticle-num-names}

\bibliography{TPMsimpleshapes1}

\end{document}